\documentclass[a4paper,titlepage,11pt]{article}

\usepackage[colorlinks=true,urlcolor=blue,citecolor=magenta]{hyperref}

\usepackage[english]{babel}

\usepackage{amsmath}
\usepackage{amssymb}
\usepackage{amsfonts}
\usepackage{amscd}
\usepackage{amsthm}
\usepackage{amsbsy}
\usepackage{latexsym}
\usepackage{bbm}

\usepackage{epsfig}
\usepackage{graphicx}
\usepackage{epstopdf}
\usepackage{psfrag}

\numberwithin{equation}{section}

\begin{document}

\begin{titlepage}

\phantom{a}
 \vskip 2.0 cm

\centerline{\Huge \bf Particle Number and} \vskip .2cm  \centerline{\Huge \bf 3D Schroedinger Holography}

\vskip 1.2cm

\centerline{\large {{\bf Jelle Hartong$^a$, Blaise Rollier$^b$}}}

\vspace{.8cm}

\begin{center}
$^a${\small\slshape Niels Bohr Institute,\\
University of Copenhagen,\\
Blegdamsvej 17, DK-2100 Copenhagen \O,
Denmark}
\vskip .2cm
$^b${\small\slshape  Institute for Theoretical Physics,\\
University of Amsterdam,\\
Science Park 904, Postbus 94485, 1090 GL Amsterdam, The Netherlands}

 \vspace{.7cm}

\centerline{{\small \texttt{hartong@nbi.dk, B.R.Rollier@uva.nl}}}

 \vspace{.3cm}
\today

\end{center}

\vskip 1cm \centerline{\bf Abstract} \vskip 0.2cm \noindent

We define a class of space-times that we call asymptotically locally Schr\"odinger space-times. We consider these space-times in 3 dimensions, in which case they are also known as null warped AdS$_3$. The boundary conditions are formulated in terms of a specific frame field decomposition of the metric which contains two parts: an asymptotically locally AdS metric and a product of a lightlike frame field with itself. Asymptotically we say that the lightlike frame field is proportional to the particle number generator $N$ regardless of whether $N$ is an asymptotic Killing vector or not.

We consider 3-dimensional AlSch space-times that are solutions of the massive vector model. We show that there is no universal Fefferman--Graham (FG) type expansion for the most general solution to the equations of motion. We show that this is intimately connected with the special role played by particle number. Fefferman--Graham type expansions are recovered if we supplement the equations of motion with suitably chosen constraints. We consider three examples. 1). The massive vector field is null everywhere. The solution in this case is exact as the FG series terminates and has $N$ as a null Killing vector. 2). $N$ is a Killing vector (but not necessarily null). 3). $N$ is null everywhere (but not necessarily Killing). The latter case contains the first examples of solutions that break particle number, either on the boundary directly or only in the bulk. Finally, we comment on the implications for the problem of holographic renormalization for asymptotically locally Schr\"odinger space-times.

\end{titlepage}

\newpage

\begingroup
\hypersetup{linkcolor=black}
\tableofcontents
\endgroup

\newpage

\section{Introduction}

Schr\"odinger space-times are relevant for at least three different reasons. First of all they are relevant for black hole physics via the Kerr/CFT correspondence \cite{Guica:2008mu}. In this context 3-dimensional Schr\"odinger space-times are often referred to as null warped AdS$_3$. The null warped AdS$_3$ space-times can be obtained as a scaling limit of space-like warped AdS$_3$ \cite{ElShowk:2011cm} which in turn plays an important role in the Kerr/CFT correspondence as well as in topologically massive gravity \cite{Anninos:2008fx}. For related work on null warped AdS$_3$ see \cite{Anninos:2010pm,Guica:2010sw,ElShowk:2011cm,Song:2011sr,Guica:2011ia}.

Secondly, Schr\"odinger space-times are generic solutions of supergravity \cite{Kraus:2011pf} and when the dynamical critical exponent $z=2$ they can be obtained from an AdS compactification via a TsT transformation or a Null Melvin twist \cite{Maldacena:2008wh,Herzog:2008wg,Adams:2008wt,Imeroni:2009cs,Hartong:2010ec,Bobev:2011qx,Detournay:2012dz}. See also \cite{Hartnoll:2008rs,Mazzucato:2008tr,Donos:2009en,Bobev:2009mw,Donos:2009xc,Colgain:2010rg} for ways of obtaining Schr\"odinger space-times in supergravity.

Thirdly, they feature in the context of potential applications of holography to strongly coupled systems with Schr\"odinger symmetries. In this setting Schr\"odinger space-times were introduced\footnote{Schr\"odinger space-times also appear in the work of \cite{Duval:2008jg}.} in \cite{Son:2008ye,Balasubramanian:2008dm} in an attempt to holographically describe strongly coupled systems with Schr\"odinger symmetries such as observed in ultracold dilute gases of point-like interacting spin-1/2 fermions fine-tuned to infinite scattering length \cite{Mehen:1999aa}. This has proven to be difficult because a Schr\"odinger space-time has particle number as one of its isometries while the groundstate of these ultracold gases are superfluids, meaning that particle number is spontaneously broken. Hence one would have to resort at least to asymptotically Schr\"odinger space-times that break particle number in the IR\footnote{See \cite{Balasubramanian:2010uw} for other ways of realizing the Schr\"odinger symmetry and breaking particle number.}. Nevertheless there are other strongly coupled Schr\"odinger invariant systems see e.g. \cite{Henkel:1993sg,Henkel:2002vd} for which holography may be of use (see for example \cite{Minic:2008xa} in the context of ageing systems, see also \cite{Leiva:2003kd} and \cite{Wang:2013tv}).

On top of these three reasons there is the additional motivation to learn about holography beyond the familiar AdS/CFT context. In this light since Schr\"odinger space-times are not asymptotically locally AdS (AlAdS) as defined in \cite{deHaro:2000xn,Papadimitriou:2005ii}, while they have at the same time a well defined relation to AlAdS space-times they provide a nice case for developing geometric tools to perform basic holographic calculations beyond the familiar AdS/CFT context. We will see that on the one hand the Schr\"odinger asymptotics provide a mild and on the other hand significant departure from standard AdS/CFT lore, much more so than in the case of asymptotically locally Lifshitz space-times (AlLif) \cite{Ross:2009ar,Ross:2011gu,Mann:2011hg,Griffin:2011xs,Baggio:2011cp,Baggio:2011ha,Chemissany:2012du}. It therefore provides a great opportunity to learn more about holography for non-AdS space-times.

To understand better that asymptotically locally $z=2$ Schr\"odinger space-times form a more significant departure from the AlAdS context than for example $z=2$ AlLif space-time we recall that the latter can be uplifted to a certain class of AlAdS space-times that asymptote to a $z=0$ Schr\"odinger space-time in one dimension higher\footnote{These observations can be extended to uplift hyperscaling violating Lifshitz space-times to one dimension higher where they become (conformal) Schr\"odinger space-times with $z<1$ obeying the null energy condition \cite{Chemissany:2011mb,Narayan:2012hk,Perlmutter:2012he,Kim:2012nb,Gath:2012pg}.}\cite{Balasubramanian:2010uk,Donos:2010tu,Costa:2010cn,Cassani:2011sv,Chemissany:2011mb,Chemissany:2012du,Narayan:2012wn}. This uplift applies to the full class of $z=2$ AlLif space-times. In contrast $z=2$ AlSch can be related via TsT transformations to AlAdS space-times of the same dimensionality but this does not apply to the full class of $z=2$ AlSch space-times. FG expansions for AlSch solutions obtained via TsT were discussed for the first time in 5-dimensions in \cite{Hartong:2010ec}. Inspired by \cite{Hartong:2010ec} we made more precise the notion of a Schr\"odinger boundary for locally Schr\"odinger space-times in \cite{Hartong:2012sw}. The current paper extends this work to the full class of AlSch space-times in 3 dimensions.

The success of the supergravity approximation of the AdS/CFT correspondence depends to a large extent on the celebrated Fefferman--Graham (FG) theorem \cite{FeffermanGraham} which in AdS/CFT language states that: given certain regularity assumptions the most general solution to the Einstein equations (possibly with matter) with AlAdS boundary conditions can be written in the form of an asymptotic expansion that depends on two sets of free functions: sources and vevs.

When we try to generalize geometric notions of the AdS/CFT correspondence to non-AdS spaces an important question is whether there is an analogue (possibly modified) version of the FG result. For AlLif space-times there is such a result, see e.g. \cite{Ross:2011gu,Chemissany:2012du}. The extent to which we have a similar result for asymptotically locally Schr\"odinger space-times (AlSch), to be defined shortly, will be the main question studied in this paper.

We now proceed to summarize our main results. We consider a model in which we have an Einstein--Hilbert term with a negative cosmological constant coupled to a massive vector field $A^\mu$. We define a class of 3-dimensional space-times that we call AlSch (in some analogy with AlLif) as follows. As discussed in section \ref{sec:AlSch} any AlSch can be written in a frame field basis as follows
\begin{equation}
g_{\mu\nu}dx^\mu dx^\nu=\left(-e^{+}_a e^{+}_b-e^{+}_a e^{-}_b-e^{+}_b e^{-}_a\right)dx^a dx^b+\frac{dr^2}{r^2}\,,
\end{equation}
where $e^{+}_a=r^{-2}\left(e^+_{(0)a}+o(1)\right)$ is a null vector normalized such that $e^{+}_a e^{-a}=-1$ with $e^{-}_a=e^{-}_{(0)a}+o(1)$ a unit spacelike vector. The vector field $A^a$ is required to asymptote to $e^{+a}$ in a manner defined precisely in section \ref{sec:AlSch} while $A^r=o(r)$.

We are now in a position to be more precise about what we mean by particle number. For a Schr\"odinger space-time particle number is a symmetry whose generator is given by the unique hypersurface orthogonal null Killing vector of the Schr\"odinger algebra \cite{Blau:2010fh}. In the solution for a Schr\"odinger space-time this is described by $A^\mu$. Whenever we are dealing with an AlSch space-time that is such that the boundary value of $A^\mu$, that we denote by $A_{(0)}^a$, is proportional to a null Killing vector we say that the UV of the boundary theory has particle number as a symmetry. However we will not demand that there should be any asymptotic symmetries present in the solution and hence we will encounter cases where $A_{(0)}^a$ is not proportional to a Killing vector. In those cases we will still say that $A_{(0)}^a$ is proportional to the particle number generator for lack of a better name.

For the purpose of solving locally the equations of motion of the massive vector model we define a special gauge, that we call the radial $TV$ gauge in which the particle number generator is $N=\partial_V$ while the Hamiltonian is $\partial_T$. In this radial $TV$ gauge the AlSch boundary conditions read
\begin{eqnarray}
e^+_a & = & r^{-2}\left(A_{(0)T}\delta^T_a+o(1)\right)\,,\\
e^-_a & = & N_{(0)V}\delta^V_a+o(1)\,,\\
A^a & = & A^V_{(0)}\delta_V^a+o(1)\,,\\
A^r & = & o(r)\,.
\end{eqnarray}
The $TV$ gauge is a choice of boundary coordinates that can be made without loss of generality. Our results do not depend on this specific set of coordinates and they can be easily written down in other coordinate systems.

It has been discussed a lot in the literature whether or not one should compactify the direction $V$ generated by particle number so that its eigenvalues can only form a discrete set. This has the obvious drawback of creating closed causal curves and circles that become asymptotically null which may lead to a breakdown of various approximations. However, not compactifying has the obvious drawback of not being able to have a discrete spectrum. There are many related issues that we will not address further here in the introduction. We just bring this up to make clear that we will be largely agnostic to this issue. Since not imposing any conditions on $V$ leads to the largest set of solutions we will treat $V$ like any other non-compact coordinate unless specifically stated otherwise. We will discuss the effect of compactifying $V$ at various instances.

The main advantage of using the radial $TV$ gauge is that we can easily discuss one of our main conclusions:
\begin{quote}
The FG theorem breaks down for solutions $g_{\mu\nu}$ and $A^\mu$ whose $V$-dependence is unconstrained.
\end{quote}
By this we mean that there is no unique radial expansion for solutions whose $V$-dependence is unconstrained. We will make this statement more precise in section \ref{sec:solutions}.

So far in the literature two types of scenarios relating to the notion of AlSch space-times have been studied:
\begin{itemize}
\item Linearized perturbations around a fixed Schr\"odinger background \cite{Guica:2010sw,vanRees:2012cw}.
\item Constrained solutions: in all cases that have been considered these turn out to be $V$-independent solutions.
\end{itemize}

We will not study linearized perturbations but the simpler case of a real massless scalar field on a fixed Schr\"odinger background. This case is studied in appendix \ref{app:scalars}. There we will see the same behavior as for solutions to the equations of motion of the massive vector model. There is no FG-like expansion for the most general solution to the Klein--Gordon equation on a fixed Schr\"odinger background.

In \cite{Guica:2011ia} the constrains $g_{\mu\nu}A^\mu A^\nu=0$ and $F_{\mu\nu}=-2\epsilon_{\mu\nu\rho}A^\rho$ have been imposed with $F_{\mu\nu}=\partial_\mu A_\nu-\partial_\nu A_\mu$ and $\epsilon_{\mu\nu\rho}$ the 3-dimensional Levi-Civita tensor. Here we will show that the only AlSch space-times satisfying these two conditions take the same form as a locally Schr\"odinger space-time which is defined in appendix \ref{app:locallySch} with the only difference being that $A^\mu$ must be proportional to the null Killing vector $(\partial_V)^\mu$ whereas for a locally Schr\"odinger space-time it is equal to the null Killing vector $(\partial_V)^\mu$. This will be shown in sections \ref{subsec:constrained} and \ref{subsubsec:A^2=0again}. There we will also show that if we only impose the $g_{\mu\nu}A^\mu A^\nu=0$ constraint the solution still does not depend on $V$ with $A^\mu$ proportional to the null Killing vector $(\partial_V)^\mu$. The FG expansion for this solution terminates making the solution an exact solution of the equations of motion. One could say that all these constrained solutions are of the TsT form by which we mean that they admit a bulk Killing vector that is null everywhere or becomes null asymptotically. The most general solution obtained by demanding $V$-independence is constructed in section \ref{subsubsec:Vindependentsolutions}. We will also discuss constrained solutions obtained by demanding that $\partial_V$ is null everywhere that have a non-trivial $V$-dependence and for which a unique $r$-expansion can be written down. These are the first examples of AlSch space-times without any Killing vectors. These come in two flavors: 1). with a flat boundary metric (seen from the AlAdS perspective) so that particle number is an asymptotic Killing vector and 2). with a non flat boundary metric (seen from the AlAdS perspective) that breaks particle number. These solutions are discussed in section \ref{subsec:moreconstrained}.

For all the asymptotic expansions constructed in sections \ref{sec:solutions} and \ref{subsec:moreconstrained} we compute in section \ref{sec:HR} the on-shell action. We show that for all constrained solutions the usual AdS$_3$ counterterms suffice but that in the case of the more general solutions of section \ref{sec:solutions} where the $V$-dependence is only constrained by our ansatz for the radial expansion the on-shell action remains logarithmically divergent, i.e. we did not find a local counterterm (possibly multiplied by $\log r$) that can remove this divergence.

This paper is organized as follows. After a brief description of our model in section \ref{subsec:model} we start by introducing the boundary conditions for AlSch space-times in section \ref{subsec:Schbdry}. After this we discuss the Schr\"odinger boundary in section \ref{subsec:definingfunction} in terms of a defining function. As a first example of AlSch space-times we first deal with the constraint case in which $A^\mu$ is required to be null in section \ref{subsec:constrained}. In section \ref{subsec:ansatz} we then make a very general ansatz and construct the asymptotic expansion up to the order that is relevant for studying the divergences of the on-shell action without imposing any constraints other than the ansatz itself. We then show the appearance of undetermined functions at each order in the radial expansion in section \ref{subsec:higher} and we comment on the dependence of the solution on the ansatz in section \ref{subsec:commentsansatz}. In particular we show that one can add arbitrarily high powers of logarithmic terms of the radial coordinate. This shows that there is no FG expansion that covers all solutions. We then proceed to show that these problems disappear when we impose suitable constraints on the solution. To this end we discuss three different constrained solutions in section \ref{subsec:moreconstrained} including a rederivation of the case where $A^\mu$ is null in the radial $TV$ gauge for the AlSch metric as well as the most general $V$-independent solution and an example that breaks particle number obtained by demanding that $\partial_V$ is null. We briefly comment on the on-shell action and counterterms in section \ref{sec:HR} for all the solutions constructed here and we end with a discussion and outlook in \ref{sec:discussion}. Finally in appendix \ref{app:locallySch} we review the case of a locally Schr\"odinger space-time, in appendix \ref{app:solutions} we collect some background information on how to construct the asymptotic expansions and in appendix \ref{app:scalars} we discuss the breakdown of the FG result for a real scalar on a fixed Schr\"odinger background.

\section{Asymptotically locally Schr\"odinger space-times}\label{sec:AlSch}

\subsection{The model}\label{subsec:model}

Three-dimensional Schr\"odinger space-times are solutions of topologically massive gravity \cite{Anninos:2008fx}, certain consistent truncations of type IIB supergravity reduced to 3 dimensions \cite{Detournay:2012dz} and of the toy model consisting of gravity coupled to a massive vector field \cite{Son:2008ye,Balasubramanian:2008dm}. Here we will consider, largely for simplicity, the massive vector model. The action we will work with
reads
\begin{equation}\label{eq:action-model}
S=\int d^{3}x\sqrt{-g}\left(R+2-\frac{1}{4}F_{\mu\nu}F^{\mu\nu}-2A_\mu A^\mu\right)\,,
\end{equation}
where $F_{\mu\nu}=\partial_\mu A_\nu-\partial_\nu A_\mu$. The equations of motion coming from \eqref{eq:action-model} are
\begin{eqnarray}
R_{\mu\nu} & = & -2g_{\mu\nu}+\frac{1}{2}F_{\mu\rho}F_\nu{}^\rho-\frac{1}{4}F_{\rho\sigma}F^{\rho\sigma}g_{\mu\nu}+2A_\mu A_\nu\,,\label{eq:eombarg}\\
\nabla_\mu F^{\mu\nu} & = & 4A^\nu\,,\label{eq:eomMatbarA}
\end{eqnarray}
and as a consequence of \eqref{eq:eomMatbarA} we have
\begin{equation}\label{eq:divA}
\nabla_\mu A^\mu=0\,.
\end{equation}

\subsection{AlSch space-times}\label{subsec:Schbdry}

We will next formulate the boundary conditions with which we are going to solve the equations of motion of the massive vector model. These boundary conditions lead to  the notion of asymptotically locally Schr\"odinger space-times.

Let $\bar{\mathcal{M}}$ be a $(d+3)$-dimensional manifold with a boundary $\partial\mathcal{M}$ and interior $\mathcal{M}$. Let there be a non-degenerate metric $\gamma'_{\mu\nu}$ that is regular on $\bar{\mathcal{M}}$. Let there also be a null vector field $e^{+\mu}$ ($\gamma_{\mu\nu}e^{+\mu} e^{+\nu}=0$) that is regular on $\bar{\mathcal{M}}$ and that does not vanish on $\partial\mathcal{M}$. Furthermore, let there be a function $\Omega$ (known as the defining function) that is zero at the boundary $\partial\mathcal{M}$ with the property that $\partial_\mu\Omega$ is nonzero at the locus $\Omega=0$. Following Penrose \cite{Penrose:1986ca} the metric $\gamma'_{\mu\nu}$ is a conformal completion of the metric $\gamma_{\mu\nu}$ defined on $\mathcal{M}$ if the two are related via $\gamma'_{\mu\nu}=\Omega^2\gamma_{\mu\nu}$. Now let there be a different metric defined on $\mathcal{M}$ that is denoted by $g_{\mu\nu}$. We say that $g_{\mu\nu}$ admits an anisotropic conformal completion \cite{Horava:2009vy} $\gamma'_{\mu\nu}-\gamma'_{\mu\rho}e^{+\rho} \gamma'_{\nu\sigma}e^{+\sigma}$ with critical exponent $z=2$ when $g_{\mu\nu}=\Omega^{-2}\gamma'_{\mu\nu}-\Omega^{-4}\gamma'_{\mu\rho}e^{+\rho} \gamma'_{\nu\sigma}e^{+\sigma}$. We have
\begin{equation}
e^+_\mu=g_{\mu\nu}e^{+\nu}=\gamma_{\mu\nu}e^{+\nu}\,.
\end{equation}
The inverse and determinant of $g_{\mu\nu}=\gamma_{\mu\nu}-e^+_\mu e^+_\nu$ are given by
\begin{eqnarray}
g^{\mu\nu} & = & \gamma^{\mu\nu}+e^{+\mu} e^{+\nu}\,,\\
\text{det}\,g_{\mu\nu} & = & \text{det}\, \gamma_{\mu\nu}\,.\label{eq:determinant}
\end{eqnarray}

We define a 3-dimensional asymptotically locally Schr\"odinger (AlSch) metric as any metric $g_{\mu\nu}$ which is such that $g_{\mu\nu}+e^+_\mu e^+_\nu$ is AlAdS with $e^{+\mu}$ a null vector of the AlAdS space-time that is asymptotically nonzero, i.e. does not vanish on $\partial\mathcal{M}$. Any 3-dimensional AlSch can be written in a frame field basis as follows
\begin{equation}\label{eq:frame fielddecompositionmetricg}
g_{\mu\nu}=-e^{+}_\mu e^{+}_\nu-e^{+}_\mu e^{-}_\nu-e^{-}_\mu e^{+}_\nu+e^{2}_\mu e^{2}_\nu\,,
\end{equation}
with
\begin{eqnarray}
e^{+}_\mu & = & \Omega^{-2}e^{+}_{(0)\mu}+\ldots\,,\label{eq:e^+}\\
e^{-}_\mu & = & e^{-}_{(0)\mu}+\ldots\,,\label{eq:e^-}\\
e^{2}_\mu & = & \Omega^{-1}e^{2}_{(0)\mu}+\ldots\,.
\end{eqnarray}
where $\Omega$ is the defining function introduced at the beginning of this subsection and where the dots denote terms that are subleading\footnote{For the sake of comparison we mention that any 3-dimensional AlLif space-time with $z=2$ can be written in a frame field basis as
\begin{equation}
g_{\mu\nu}=-e^{0}_\mu e^{0}_\nu+e^{1}_\mu e^{1}_\nu+e^{2}_\mu e^{2}_\nu\,,
\end{equation}
where
\begin{eqnarray}
e^{0}_\mu & = & \Omega^{-2}e^{0}_{(0)\mu}+\ldots\,,\\
e^{1}_\mu & = & \Omega^{-1}e^{1}_{(0)\mu}+\ldots\,,\\
e^{2}_\mu & = & \Omega^{-1}e^{2}_{(0)\mu}+\ldots\,.
\end{eqnarray}
}. The frame fields satisfy
\begin{eqnarray}
e^{+}_\mu e^{+\mu} & = & 0\,,\label{eq:frame field1}\\
e^{+}_\mu e^{-\mu} & = & -1\,,\\
e^{-}_\mu e^{-\mu} & = & 1\,,\\
e^{+}_\mu e^{2\mu} & = & 0\,,\\
e^{-}_\mu e^{2\mu} & = & 0\,,\\
e^{2}_\mu e^{2\mu} & = & 1\,,\label{eq:frame field6}
\end{eqnarray}
so that the expansion of their inverses reads
\begin{eqnarray}
e^{+\mu} & = & e^{+\mu}_{(0)}+\ldots\,,\\
e^{-\mu} & = & -e^{+\mu}_{(0)}+\ldots\,,\\
e^{+\mu}+e^{-\mu} & = & \Omega^2e^{-\mu}_{(0)}+\ldots\,,\label{eq:bar e under V}\\
e^{2\mu} & = & \Omega e^{2\mu}_{(0)}+\ldots\,,
\end{eqnarray}
with
\begin{eqnarray}
e^{+}_{(0)\mu}e^{+\mu}_{(0)} & = & 0\,,\\
e^{-}_{(0)\mu}e^{+\mu}_{(0)} & = & -1\,,\\
e^{-}_{(0)\mu}e^{-\mu}_{(0)} & = & 0\,.
\end{eqnarray}
The boundary metric of the AlAdS$_3$ space-time $g_{\mu\nu}+e^{+}_\mu e^{+}_\nu$ is
\begin{equation}
\gamma_{(0)ab}=-e^{+}_{(0)a}e^{-}_{(0)b}-e^{+}_{(0)b}e^{-}_{(0)a}\,.
\end{equation}
The $\Omega^2$ term in \eqref{eq:bar e under V} can also be obtained by demanding that $\gamma_{\mu\nu}=g_{\mu\nu}+e^{+}_\mu e^{+}_\nu$ is AlAdS$_3$ implying that the inverse metric
\begin{equation}\label{eq:frame fieldinvmetric}
\gamma^{\mu\nu}=g^{\mu\nu}-e^{+\mu}e^{+\nu}=-e^{+\mu}\left(e^{+\nu}+e^{-\nu}\right)-e^{+\nu}\left(e^{+\mu}+e^{-\mu}\right)+e^{2\mu}e^{2\nu}
\end{equation}
is of order $\Omega^2$. The AlAdS$_3$ metric $g_{\mu\nu}+e^{+}_\mu e^{+}_\nu$ has the frame field decomposition $-e^{+}_\mu e^{-}_\nu-e^{-}_\mu e^{+}_\nu+e^{2}_\mu e^{2}_\nu$ but we note that the inverse frame fields obtained by raising the space-time index with the AlAdS$_3$ metric $\gamma_{\mu\nu}$ satisfies different relations than \eqref{eq:frame field1}--\eqref{eq:frame field6}. The inverse frame fields for the AlAdS$_3$ metric are given in \eqref{eq:frame fieldinvmetric}.

In the case of the massive vector model we furthermore need a boundary condition for $A^\mu$ which has the following frame field decomposition
\begin{equation}\label{eq:frame fielddecompositionA}
A^\mu=A_{+}e^{+\mu}+A_{-}e^{-\mu}+A_{2}e^{2\mu}\,.
\end{equation}
We demand that $A^\mu$ asymptotes to $e^{+\mu}$. More precisely by this we mean that
\begin{eqnarray}
A^\mu e_\mu^{+}\vert_{\Omega=0} & = & 0\,,\\
A^\mu e_\mu^{-}\vert_{\Omega=0} & = & -1\,,\\
A^\mu e_\mu^{2}\vert_{\Omega=0} & = & 0\,.\label{eq:Awithe2}
\end{eqnarray}
This means that
\begin{eqnarray}
A_{+}\vert_{\Omega=0} & = & 1\,,\label{eq:A+}\\
A_{-}\vert_{\Omega=0} & = & 0\,,\label{eq:A-}\\
A_{2}\vert_{\Omega=0} & = & 0\,.\label{eq:A2}
\end{eqnarray}

\subsection{The defining function}\label{subsec:definingfunction}

We will now take a closer look at the defining function and define what we mean by a Schr\"odinger defining function. This will allow us to define AlSch boundary conditions in radial gauge.

By taking $g_{\mu\nu}=\gamma_{\mu\nu}-e^+_\mu e^+_\nu$ and conformally rescaling $\gamma_{\mu\nu}=\Omega^{-2}\gamma'_{\mu\nu}$ we get for \eqref{eq:divA}
\begin{equation}
\nabla^{(\gamma')}_\mu A^\mu-3A^\mu\Omega^{-1}\partial_\mu\Omega=0\,.
\end{equation}
Using
\begin{equation}
\left(\nabla^{(\gamma')}_\mu A^\mu-3\Omega^{-1}A^\mu\partial_\mu\Omega\right)\vert_{\Omega=0}=0\,,
\end{equation}
we conclude that near $\Omega=0$
\begin{equation}\label{eq:normalexpA}
A^\mu\partial_\mu\Omega=\text{at most of order $\Omega$}.
\end{equation}
Since $\Omega$ is a defining function for an AlAdS space-time we also have by definition
\begin{equation}\label{eq:AlAdSbdrycondition}
\Omega^4\left(R^{(\gamma)}_{\mu\nu\rho\sigma}+\gamma_{\mu\rho}\gamma_{\nu\sigma}-\gamma_{\mu\sigma}\gamma_{\nu\rho}\right)\vert_{\Omega=0}=0\,,
\end{equation}
so that (see e.g. \cite{Graham:1999jg})
\begin{equation}
\gamma^{\mu\nu}\Omega^{-2}\partial_\mu\Omega\partial_\nu\Omega\vert_{\Omega=0}=1\,.
\end{equation}
It follows that
\begin{equation}
g^{\mu\nu}\Omega^{-2}\partial_\mu\Omega\partial_\nu\Omega\vert_{\Omega=0}=1+\left(e^{+\mu} \Omega^{-1}\partial_\mu\Omega\right)^2\vert_{\Omega=0}\,,
\end{equation}
so that the boundary at $\Omega=0$ is also timelike with respect to the AlSch metric. Using \eqref{eq:frame fielddecompositionA} and \eqref{eq:A+}--\eqref{eq:A2} we have
\begin{equation}
A^\mu\Omega^{-1}\partial_\mu\Omega\vert_{\Omega=0}=e^{+\mu}\Omega^{-1}\partial_\mu\Omega\vert_{\Omega=0}\,.
\end{equation}

We will now show that the AlSch boundary conditions of the previous subsection imply that
\begin{equation}\label{eq:Schdefiningfunction}
A^\mu\Omega^{-1}\partial_\mu\Omega\vert_{\Omega=0}=r^{-1}A^r\vert_{r=0}=0\,.
\end{equation}
To this end we will employ radial gauge for the AlSch metric which in terms of the frame fields means that we take
\begin{equation}\label{eq:radialgauge}
e^+_r=e^-_r=e^2_a=0\,,\qquad e^2_r=\frac{1}{r}\,.
\end{equation}
In radial gauge equation \eqref{eq:Awithe2} tells us that
\begin{equation}\label{eq:Schdefiningfunctionradial}
\frac{1}{r}A^r\vert_{r=0}=0\,.
\end{equation}
Further in radial gauge we can without loss of generality take $\Omega=r$ so that \eqref{eq:Schdefiningfunction} becomes \eqref{eq:Schdefiningfunctionradial}. Hence the AlSch boundary conditions enforce \eqref{eq:Schdefiningfunction}.

Taking the AlSch metric $g_{\mu\nu}$ in radial gauge implies that the AlAdS metric $\gamma_{\mu\nu}=g_{\mu\nu}+e^+_\mu e^+_\nu$ is also in radial gauge. Even if we had not imposed the boundary condition \eqref{eq:Awithe2} we could without loss of generality have chosen a gauge such that \eqref{eq:Schdefiningfunctionradial} holds. To see this we need to show that among the radial gauge preserving diffeomorphisms there is a transformation that sets $A^r$ equal to zero at leading order. To this end consider the AlAdS metric in radial gauge
\begin{equation}\label{eq:AlAdSradialgauge}
ds^2=\gamma_{\mu\nu}dx^\mu dx^\nu=\frac{dr^2}{r^2}+\gamma_{ab}dx^a dx^b\,.
\end{equation}
This radial gauge is preserved by diffeomorphisms that are generated by $\delta \gamma_{\mu\nu}=\nabla_\mu\xi_\nu+\nabla_\nu\xi_\mu$ such that $\delta \gamma_{rr}=\delta \gamma_{ra}=0$. These are given by
\begin{eqnarray}
\xi^r & = & r\xi^r_{(0)}\,,\label{eq:xi-r}\\
\xi^a & = & \xi^a_{(0)}-\int \frac{dr}{r}\gamma^{ab}\partial_b\xi^r_{(0)}\,.\label{eq:xi-a}
\end{eqnarray}
These are also known as Penrose--Brown--Henneaux (PBH) transformations \cite{Penrose:1986ca,Brown:1986nw}. Under these PBH transformations the $r$-component of $A_\mu$ transforms as
\begin{equation}
\delta A_r  =  \xi^\mu\partial_\mu A_r+A_\mu\partial_r\xi^\mu=r\xi_{(0)}^r\partial_rA_r+\xi^a\partial_aA_r+\xi^r_{(0)}A_r-\frac{1}{r}A^a\partial_a\xi_{(0)}^r\,.
\end{equation}
Using \eqref{eq:normalexpA} together with the fact that in radial gauge we can take $\Omega=r$ so that
\begin{equation}
A^r=rA_{(0)}^r+\ldots\,,
\end{equation}
implying that
\begin{equation}
A_r=\frac{1}{r}A_{(0)r}+\ldots\,,
\end{equation}
where the dots denote subleading terms. Using this expansion together with the fact that we have by definition (recall that $A^\mu$ is non-vanishing on the boundary)
\begin{equation}
A^a=A^a_{(0)}+\ldots
\end{equation}
we get for the transformation of the leading term in the expansion of $A_r$
\begin{equation}
A'_{(0)r}=A_{(0)r}-\delta A_{(0)r}=A_{(0)r}-\xi_{(0)}^a\partial_aA_{(0)r}+A_{(0)}^a\partial_a\xi_{(0)}^r\,.
\end{equation}
We can thus choose a gauge, on top of the radial gauge choice, such that $A'_{(0)r}=0$. Hence we can always choose coordinates on an AlAdS space-time such that \eqref{eq:Schdefiningfunction} as well as all the other usual requirements for $\Omega$ to be a defining function are satisfied. This gauge is preserved by all $\xi^\mu$ of the form
\eqref{eq:xi-r} and \eqref{eq:xi-a} where we additionally need $A_{(0)}^a\partial_a\xi_{(0)}^r=0${\,}\footnote{Such diffeomorphisms have been considered previously in \cite{Hartong:2010ec}.}. This is a gauge in which the boundary at $\Omega=0$ has been oriented such that $A^\mu$ is tangential to it. We will call $\Omega$ a Schr\"odinger defining function when it also (on top of the earlier requirements for it to be an AdS defining function) satisfies \eqref{eq:Schdefiningfunction} (see also section \ref{subsec:boundary} of appendix \ref{app:locallySch}).

We conclude that in radial gauge the AlSch boundary conditions of the previous subsection read
\begin{equation}\label{eq:radialgaugeframe fielddecomposition}
g_{\mu\nu}dx^\mu dx^\nu=\left(-e^{+}_a e^{+}_b-e^{+}_a e^{-}_b-e^{+}_b e^{-}_a\right)dx^a dx^b+\frac{dr^2}{r^2}\,,
\end{equation}
\begin{eqnarray}
e^+_a & = & r^{-2}\left(e^+_{(0)a}+o(1)\right)\,,\label{eq:e+}\\
e^-_a & = & e^-_{(0)a}+o(1)\,,\\
A^a & = & e^{+a}_{(0)}+o(1)\,,\\
A^r & = & o(r)\,.\label{eq:Ar}
\end{eqnarray}

\subsection{Local Lorentz transformations seen from the boundary}

In the beginning of section \ref{subsec:Schbdry} we said that $g_{\mu\nu}=\Omega^{-2}\gamma'_{\mu\nu}-\Omega^{-4}\gamma'_{\mu\rho}e^{+\rho} \gamma'_{\nu\sigma}e^{+\sigma}$ admits an anisotropic conformal completion $\gamma'_{\mu\nu}-\gamma'_{\mu\rho}e^{+\rho} \gamma'_{\nu\sigma}e^{+\sigma}$. Evaluating this on the boundary we obtain the metric $\left(\gamma'_{ab}-\Omega^4e^+_ae^+_b\right)\vert_{\Omega=0}=-e^+_{(0)a}e^-_{(0)b}-e^+_{(0)b}e^-_{(0)a}-e^+_{(0)a}e^+_{(0)b}$. We will nevertheless not call this the `Schr\"odinger boundary metric'. In fact we will show that the bulk local Lorentz transformations acting on $e^{\underline{a}}_a$ where $\underline{a}=-,+$ of the tangent space of the constant $r$ slices which is $SO(1,1)$ induces an action on the leading components of the frame fields that is not of the form of a Lorentz transformation but rather takes the form of a Galilean boost. We therefore refrain from building Lorentzian boundary metrics in terms of $e^{\underline{a}}_{(0)a}$.

To find the action of a bulk local Lorentz transformation on the boundary fields $e_{(0)a}^+$ and $e_{(0)a}^-$ we proceed as follows. Consider the bulk $SO(1,2)$ Lorentz transformations that leave $e^2_\mu=r^{-1}\delta_\mu^r$ inert. This leaves us with the following $SO(1,1)$ transformation
\begin{eqnarray}
e_{a}^+ & = & \lambda e_{a}'^+\,,\\
e_{a}^- & = & \frac{1}{\lambda} e_{a}'^-+\frac{1}{2}\left(\frac{1}{\lambda}-\lambda\right)e_{a}'^+\,,
\end{eqnarray}
where $\lambda$ is a function of the bulk space-time coordinates. Expanding the left hand side we obtain
\begin{eqnarray}
e_{a}'^+ & = & \frac{1}{\lambda}\frac{1}{r^2}e_{(0)a}^++\ldots\,,\\
e_{a}'^- & = & \lambda e_{(0)a}^-+\frac{\left(\lambda^2-1\right)}{2\lambda}\frac{1}{r^2}e_{(0)a}^++\ldots\,,
\end{eqnarray}
where we did not yet expand $\lambda$ in $r$. Demanding that this can again be written as
\begin{eqnarray}
e_{a}'^+ & = & \frac{1}{r^2}e_{(0)a}'^++\ldots\,,\\
e_{a}'^- & = & e_{(0)a}'^-+\ldots\,,
\end{eqnarray}
we conclude that we need
\begin{equation}
\lambda=1+r^2\lambda_{(0)}+\ldots\,.
\end{equation}
However this means that $e_{(0)a}^+$ and $e_{(0)a}^-$ transform as
\begin{eqnarray}
e_{(0)a}'^+ & = & e_{(0)a}^+\,,\label{eq:affine1}\\
e_{(0)a}'^- & = & e_{(0)a}^-+\lambda_{(0)}e_{(0)a}^+\,,\label{eq:affine2}
\end{eqnarray}
which is not a (1+1)-dimensional Lorentz transformation but a Galilean boost. We conclude that the bulk does not induce a Lorentzian metric structure on the boundary for which $e_{(0)a}^+$ and $e_{(0)a}^-$ are the frame fields. However certain structures do carry over from the bulk to boundary. For example the determinant of $e_{(0)a}^{\underline{a}}$ is left invariant under the transformation \eqref{eq:affine1} and \eqref{eq:affine2}. Hence, it follows that the AlSch metric turns its tangent Lorentz group into a different type of group when seen from the boundary perspective. It would be interesting to work out further the metric structures that the bulk induces on the boundary.

\section{$A^\mu$ is null: an AlAdS point of view}\label{subsec:constrained}

In the next section we will discuss various solutions to the equations of motion of the massive vector model that satisfy the boundary conditions of an AlSch space-time. Here we start for simplicity and to get an idea about the possible solutions with a simpler case obtained by imposing the on-shell constraint
\begin{equation}\label{eq:constraintA2=0}
g_{\mu\nu}A^\mu A^\nu=0\,.
\end{equation}
We show that in this case the FG expansion terminates leading to an exact solution.

We can without loss of generality take
\begin{equation}
A^\mu=e^{+\mu}\,.
\end{equation}
This allows us to perform a field redefinition from $g_{\mu\nu}$ to $\gamma_{\mu\nu}=g_{\mu\nu}+A_\mu A_\nu$ and to solve the resulting equations of motion for $\gamma_{\mu\nu}$ with AlAdS boundary conditions. This will be the approach taken in this subsection. Later in section \ref{subsec:moreconstrained} we will again consider this constraint case but from a Schr\"odinger perspective, i.e. by directly solving for $g_{\mu\nu}$.

\subsection{The equations for $\gamma_{\mu\nu}$}

The equations \eqref{eq:eombarg}--\eqref{eq:divA} written in terms of $\gamma_{\mu\nu}=g_{\mu\nu}+A_\mu A_\nu$ with $\gamma_{\mu\nu}A^\mu A^\nu=0$ read
\begin{eqnarray}
G^{(\gamma)}_{\mu\nu}-\gamma_{\mu\nu} & = & \frac{1}{2}\left(\mathcal{L}_AS_{\mu\nu}-S_{\mu\rho}S_\nu{}^\rho\right)+\frac{1}{4}\left(F_{\rho\sigma}F^{\rho\sigma}+2X_\rho X^\rho\right)A_\mu A_\nu-\frac{1}{8}S_{\rho\sigma}S^{\rho\sigma}\gamma_{\mu\nu}\,,\label{eq:Einsteineqs-AdSside-A2=0}\\
0 & = & \nabla^{(\gamma)}_\mu S^\mu{}_\nu+\frac{1}{2}\left(F_{\rho\sigma}F^{\rho\sigma}+2X_\rho X^\rho\right)A_\nu\,,\label{eq:vectoreq1-AdSside-A2=0}\\
0 & = & \nabla^{(\gamma)}_\mu A^\mu\,,\label{eq:vectoreq2-AdSside-A2=0}
\end{eqnarray}
where $\mathcal{L}_A$ denotes the Lie derivative along $A$ and where we defined
\begin{eqnarray}
X^\mu & = & A^\rho\nabla^{(\gamma)}_\rho A^\mu\,,\\
S_{\mu\nu} & = & \nabla^{(\gamma)}_\mu A_\nu+\nabla^{(\gamma)}_\nu A_\mu\,.
\end{eqnarray}
Indices are raised and lowered with respect to $\gamma_{\mu\nu}$.
Contracting equation \eqref{eq:vectoreq1-AdSside-A2=0} with $A^\nu$ we get
\begin{equation}\label{eq:corollary1}
\nabla^{(\gamma)}_\mu X^\mu=\frac{1}{2}S_{\mu\nu}S^{\mu\nu}\,.
\end{equation}
Contracting equation \eqref{eq:Einsteineqs-AdSside-A2=0} with $A^\mu A^\nu$ we get
\begin{equation}\label{eq:corollary2}
A^\mu A^\nu R^{(\gamma)}_{\mu\nu}=-\frac{1}{2}X_\mu X^\mu\,.
\end{equation}
Further for any null vector $A^\mu$ satisfying \eqref{eq:vectoreq2-AdSside-A2=0} we have the identity
\begin{equation}\label{eq:corollary3}
A^\mu A^\nu R^{(\gamma)}_{\mu\nu}=\nabla^{(\gamma)}_\mu X^\mu-\frac{1}{4}S_{\mu\nu}S^{\mu\nu}+\frac{1}{4}F_{\mu\nu}F^{\mu\nu}\,.
\end{equation}
Combining \eqref{eq:corollary1}--\eqref{eq:corollary3} we get
\begin{equation}\label{eq:corollary4}
F_{\mu\nu}F^{\mu\nu}+2X_\mu X^\mu=-S_{\mu\nu}S^{\mu\nu}\,.
\end{equation}
Using \eqref{eq:corollary4} we obtain the system of equations
\begin{eqnarray}
0 = \mbox{Eins}_{\mu\nu} & \equiv & R^{(\gamma)}_{\mu\nu}+2\gamma_{\mu\nu}-\frac{1}{2}\left(\mathcal{L}_AS_{\mu\nu}-S_{\mu\rho}S_\nu{}^\rho\right)+\frac{1}{4}S_{\rho\sigma}S^{\rho\sigma}A_\mu A_\nu\nonumber\\
&&-\frac{1}{4}S_{\rho\sigma}S^{\rho\sigma}\gamma_{\mu\nu}\,,\label{eq:Einsteineqs-AdSside-A2=0-v2}\\
0 =\mbox{Vec}_\nu & \equiv & \nabla^{(\gamma)}_\mu S^\mu{}_\nu-\frac{1}{2}S_{\rho\sigma}S^{\rho\sigma}A_\nu\,,\label{eq:vectoreq1-AdSside-A2=0-v2}
\end{eqnarray}
together with \eqref{eq:vectoreq2-AdSside-A2=0} and \eqref{eq:constraintA2=0}. Contracting equation \eqref{eq:Einsteineqs-AdSside-A2=0-v2} with $A^\mu$ and using the identity
\begin{equation}
R^{(\gamma)}_{\mu\nu}A^\mu=\frac{1}{2}\nabla^{(\gamma)}_\mu S^\mu{}_\nu-\frac{1}{2}\nabla^{(\gamma)}_\mu F^\mu{}_\nu
\end{equation}
which requires \eqref{eq:vectoreq2-AdSside-A2=0} as well as \eqref{eq:vectoreq1-AdSside-A2=0-v2} we obtain the following alternative vector field equation of motion
\begin{equation}
\nabla^{(\gamma)}_\mu F^{\mu\nu}+\mathcal{L}_AX^\nu=4A^\nu\,.
\end{equation}

In 3 space-time dimensions the Riemann and Einstein tensors contain the same amount of information due to the identity\footnote{We take
\begin{equation}
\epsilon_{\mu\nu\lambda}\epsilon_{\rho\sigma\kappa}=-\gamma_{\mu\rho}\gamma_{\nu\sigma}\gamma_{\lambda\kappa}+\text{permutations}\,.
\end{equation}
Further, because of the constraint $g_{\mu\nu}A^\mu A^\nu=0$, the Levi-Civita tensor is the same in terms of the metric $\gamma_{\mu\nu}$ as well as in terms of the metric $g_{\mu\nu}$.}
\begin{equation}
R^{(\gamma)}_{\mu\nu\rho\sigma}=\epsilon_{\mu\nu\lambda}\epsilon_{\rho\sigma\kappa}G^{(\gamma)\lambda\kappa}\,,
\end{equation}
where $\epsilon_{\mu\nu\lambda}$ is the 3-dimensional Levi-Civita tensor. The 3-dimensional Riemann tensor is given by
\begin{equation}
R^{(\gamma)}_{\mu\nu\rho\sigma}=\gamma_{\mu\rho}R^{(\gamma)}_{\nu\sigma}-\gamma_{\mu\sigma}R^{(\gamma)}_{\nu\rho}-\gamma_{\nu\rho}R^{(\gamma)}_{\mu\sigma}+\gamma_{\nu\sigma}R^{(\gamma)}_{\mu\rho}-\frac{1}{2}R^{(\gamma)}\left(\gamma_{\mu\rho}\gamma_{\nu\sigma}-\gamma_{\mu\sigma}\gamma_{\nu\rho}\right)\,.
\end{equation}
Using \eqref{eq:Einsteineqs-AdSside-A2=0-v2} and its trace $R^{(\gamma)}=-6+\tfrac{3}{4}S_{\rho\sigma}S^{\rho\sigma}$ we obtain
\begin{eqnarray}
R^{(\gamma)}_{\mu\nu\rho\sigma} & = & -\left(\gamma_{\mu\rho}\gamma_{\nu\sigma}-\gamma_{\mu\sigma}\gamma_{\nu\rho}\right)+\frac{1}{8}S_{\lambda\tau}S^{\lambda\tau}\left(\gamma_{\mu\rho}\gamma_{\nu\sigma}-\gamma_{\mu\sigma}\gamma_{\nu\rho}\right)\nonumber\\
&&-\frac{1}{4}S_{\lambda\tau}S^{\lambda\tau}\left(\gamma_{\mu\rho}A_\nu A_\sigma-\gamma_{\mu\sigma}A_\nu A_\rho-\gamma_{\nu\rho}A_\mu A_\sigma+\gamma_{\nu\sigma}A_\mu A_\rho\right)\nonumber\\
&&+\frac{1}{2}\left(\gamma_{\mu\rho}Y_{\nu\sigma}-\gamma_{\mu\sigma}Y_{\nu\rho}-\gamma_{\nu\rho}Y_{\mu\sigma}+\gamma_{\nu\sigma}Y_{\mu\rho}\right)\,,\label{eq:Riemann-A2=0}
\end{eqnarray}
where
\begin{equation}
Y_{\mu\nu}=\mathcal{L}_AS_{\mu\nu}-S_{\mu\rho}S_\nu{}^\rho=\gamma_{\mu\rho}\mathcal{L}_AS^\rho{}_\nu\,.
\end{equation}

\subsection{Boundary conditions}\label{subsubsec:boundaryconditions}

We want to solve the equations of motion \eqref{eq:Einsteineqs-AdSside-A2=0-v2} and \eqref{eq:vectoreq1-AdSside-A2=0-v2} as well as \eqref{eq:vectoreq2-AdSside-A2=0} and \eqref{eq:constraintA2=0} using a FG expansion for the metric $\gamma_{\mu\nu}$ and the vector field $A^\mu$. We put $\gamma_{\mu\nu}$ in a radial gauge and use the following boundary conditions
\begin{eqnarray}
\gamma_{\mu\nu}dx^\mu dx^\nu &=& \frac{dr^2}{r^2} + \gamma_{ab}dx^adx^b\,,\label{eq:ansatz1}\\
\gamma_{ab} &=& \frac{1}{r^2}\left( \gamma_{(0)ab} + \dots \right)\,,\\
A^a &=& A^a_{(0)} + \dots\,,\\
A^r &=& r\left(A^r_{(0)} + \dots\right)\,,\label{eq:ansatz4}
\end{eqnarray}
with $A^r_{(0)}=0$ (as explained in section \ref{subsec:definingfunction}) and where the dots are in principle any higher order terms that go to zero in the near boundary limit $r\rightarrow 0$.

From the leading order of the equations of motions and the $\gamma_{\mu\nu}A^\mu A^\nu=0$ constraint it follows that
\begin{eqnarray}
\nabla_{(0)a}A^a_{(0)} & = & 0\,,\\
S_{(0)ab}S_{(0)}^{ab} & = & 0\,,\\
\mathcal{L}_{A_{(0)}}S_{(0)ab}-S_{(0)ac}S_{(0)b}{}^c & = & 0\,,\label{eq:Y0ab=0}\\
\gamma_{(0)ab}A^a_{(0)}A^b_{(0)} & = & 0\,.
\end{eqnarray}
Indices are raised and lowered with respect to the boundary metric $\gamma_{(0)ab}$. Contracting \eqref{eq:Y0ab=0} with $A_{(0)}^aA_{(0)}^b$ we find
\begin{equation}
\gamma_{(0)ab}X_{(0)}^aX_{(0)}^b=0\,,
\end{equation}
where $X_{(0)}^a=A_{(0)}^b\nabla_{(0)b}A_{(0)}^a$. Since both $A^a_{(0)}$ and $X^a_{(0)}$ are null and orthogonal to each other it must be that they are proportional $X^a_{(0)} = \lambda A^a_{(0)}$. It then follows from $S_{(0)ab}S_{(0)}^{ab}=0$ that $\lambda=0$ such that we actually also have
\begin{equation}\label{eq: X is zero}
X^a_{(0)} = 0\,.
\end{equation}
Moreover, since we have a two-dimensional boundary the Einstein tensor of $\gamma_{(0)ab}$ vanishes identically and hence the equation $R_{(0)ab}A_{(0)}^aA_{(0)}^b=0$ is automatically satisfied. It then follows from $0=4R_{(0)ab}A_{(0)}^aA_{(0)}^b=F_{(0)ab}F_{(0)}^{ab}-S_{(0)ab}S^{ab}_{(0)}$ and $S_{(0)ab}S_{(0)}^{ab}=0$ that $F_{(0)ab}F_{(0)}^{ab}=0$. This in turn implies (in 2 dimensions) that
\begin{equation}\label{eq:F0ab}
F_{(0)ab}=0\,.
\end{equation}
Since $S_{(0)ab}$ is traceless and because $A_{(0)}^aS_{(0)ab}=X_{(0)b}=0$ it follows that in 2 dimensions $S_{(0)ab}$ must be proportional to $A_{(0)a}A_{(0)b}$. The condition \eqref{eq:Y0ab=0} then simplifies to $\mathcal{L}_AS_{(0)ab}=0$ or what is in this case the same $A_{(0)}^c\nabla_{(0)c}\nabla_{(0)a}A_{(0)b}=0$. This can then finally be written as
$R_{(0)abcd}A^b_{(0)}A^d_{(0)} = 0$. In two dimensions we have $R_{(0)abcd}=\tfrac{1}{2}R_{(0)}(\gamma_{(0)ac}\gamma_{(0)bd}-\gamma_{(0)ad}\gamma_{(0)bc})$ so that it follows from  $R_{(0)abcd}A^b_{(0)}A^d_{(0)} = 0$ that in fact $R_{(0)}=0$ and that therefore the boundary metric is flat.

\subsection{The solution}\label{subsec:A^2=0solution}

As shown in appendix \ref{subsection:thesolutions} we have found all solutions to the full set of equations \eqref{eq:constraintA2=0}, \eqref{eq:vectoreq2-AdSside-A2=0}, \eqref{eq:Einsteineqs-AdSside-A2=0-v2} and \eqref{eq:vectoreq1-AdSside-A2=0-v2}. They turn out to be exact solutions given by
\begin{eqnarray}
\gamma_{ab} & = & \frac{1}{r^2}\gamma_{(0)ab}+\left(a_{(2)}+\frac{2}{3}r^4\alpha_{(4)}^2\right)A_{(0)a}A_{(0)b}\,,\label{eq:solAdSsideg}\\
A^a & = & A^a_{(0)} + r^4\alpha_{(4)}A^a_{(0)}\,,\label{eq:solAdSsideA}\\
A^r & = & 0\,,\label{eq:solAdSsideAr}
\end{eqnarray}
with
\begin{eqnarray}
\nabla^{(0)}_aA_{(0)b} & = & \sigma_{(0)} A_{(0)a}A_{(0)b}\,,\qquad\mathcal{L}_{A_{(0)}}\sigma_{(0)}=0\,,\label{eq:AproptoNKV}\\
R_{(0)} & = & 0\,, \\
\mathcal{L}_{A_{(0)}}a_{(2)} & = & 0\,,  \\
\gamma_{(0)ab}A^a_{(0)}A^b_{(0)} & = & 0\,,\\
\mathcal{L}_{A_{(0)}}\alpha_{(4)} & = & 0\,.\label{eq:conditiona4}
\end{eqnarray}
We have checked that this solution solves the full equations of motion \eqref{eq:Einsteineqs-AdSside-A2=0-v2}, \eqref{eq:vectoreq1-AdSside-A2=0-v2} and constraints \eqref{eq:constraintA2=0}, \eqref{eq:vectoreq2-AdSside-A2=0}.

In order to recover our definition of a locally Schr\"odinger space-time we would need to put $\alpha_{(4)}=0$ and choose $A^a_{(0)}$ to be a null Killing vector with respect to the boundary metric $\gamma_{(0)ab}$. In particular, it does not follow from the equations of motion and constraints that $A^a_{(0)}$ is a Killing vector of $\gamma_{(0)ab}$ in the solution above as $\sigma_{(0)}$ remains an arbitrary function constrained only by $\mathcal{L}_{A_{(0)}}\sigma_{(0)}=0$. It must however be that $A^a_{(0)}$ is proportional to a null Killing vector. This follows from equation \eqref{eq:AproptoNKV}.

We can reconstruct the corresponding most general 3-dimensional AlSch solution with $g_{\mu\nu}A^\mu A^\nu=0$ via $g_{\mu\nu}=\gamma_{\mu\nu}-A_\mu A_\nu$. We find
\begin{eqnarray}
g_{\mu\nu}dx^\mu dx^\nu &=& \frac{dr^2}{r^2} + g_{ab}dx^adx^b\,,\label{eq:A^2=0}\\
g_{ab} & = & - \frac{1}{r^4}A_{(0)a}A_{(0)b} + \frac{1}{r^2}\gamma_{(0)ab}+a_{(2)}A_{(0)a}A_{(0)b}-\frac{1}{3}r^4\alpha_{(4)}^2A_{(0)a}A_{(0)b}\,,\label{eq:solAdSsideg2}\\
A^a & = & A^a_{(0)} + r^4\alpha_{(4)}A^a_{(0)}\,,\label{eq:solAdSsideA2}\\
A^r & = & 0\,,\label{eq:solAdSsideAr2}
\end{eqnarray}
where we have absorbed $-2\alpha_{(4)}$ in the arbitrariness of $a_{(2)}$. The coefficient $a_{(4)}$ is any function obeying \eqref{eq:conditiona4}. In section \ref{sec:HR} we show that one can perform holographic renormalization for this class of solutions.

In \cite{Guica:2011ia} an additional constraint is imposed (on top of $\gamma_{\mu\nu}A^\mu A^\nu=0$) which reads
\begin{equation}\label{eq:dualF}
F_{\mu\nu}=-2\epsilon_{\mu\nu\rho}A^\rho\,.
\end{equation}
This extra constraint enforces
\begin{equation}
\alpha_{(4)}=0\,.
\end{equation}
This solution is therefore identical to a locally Schr\"odinger space-time as given in \ref{eq:FGexpansions} up to the fact that $A^a_{(0)}$ is not required to be a null Killing vector of $\gamma_{(0)ab}$ but proportional to a null Killing vector\footnote{In \cite{Guica:2011ia} the asymptotic expansion is such that in \eqref{eq:ansatz4} the function $A_{(0)}^r$ is allowed to be nonzero. As shown in section \ref{subsec:definingfunction} we can turn $A_{(0)}^r$ on by a PBH transformation.}.

In \cite{Duval:2012qr} a notion of a `Schr\"odinger manifold' has been defined that in our language amounts to demanding that the AlSch metric admits a nowhere vanishing null Killing vector whose boundary value equals that of $e^{+\mu}$. In the solution \eqref{eq:A^2=0}--\eqref{eq:solAdSsideAr2} the massive vector field $A^\mu=e^{+\mu}$ is proportional to a null Killing vector field and the proportionality factor is constant on the boundary if and only if $A_{(0)}^a$ is a Killing vector of the AlAdS$_3$ boundary metric $\gamma_{(0)ab}$. Later in section \ref{subsec:moreconstrained} we will show that any AlSch space-time admitting a null Killing vector is of the form \eqref{eq:A^2=0}--\eqref{eq:solAdSsideAr2} so that all Schr\"odinger manifolds in the sense of \cite{Duval:2012qr} are described by \eqref{eq:A^2=0}--\eqref{eq:solAdSsideAr2} with the additional condition that $A_{(0)}^a$ is a Killing vector of $\gamma_{(0)ab}$. We will discuss this class of solutions further in section \ref{subsubsec:A^2=0again}.

\section{Breakdown of the Fefferman--Graham theorem for unconstrained solutions}\label{sec:solutions}

In the previous section there was a natural identification possible between $e^{+}_\mu$ and $A_\mu$. In general this is not possible as $A^\mu$ is only asymptotically null but not everywhere. In the case without constraints, which is the subject of this subsection, we will therefore work directly in terms of the AlSch metric $g_{\mu\nu}$ by making an ansatz for the subleading terms in a radial expansion normal to the Schr\"odinger boundary imposing the boundary conditions of section \ref{sec:AlSch}. This ansatz which is an expansion in terms of $r^{2n}\log^m r$ will be discussed in the next subsection. In sections \ref{subsec:propssolution} and \ref{subsection:solvingeom} we construct the solution at next to leading order (NLO) and discuss its properties. We then proceed in section \ref{subsec:higher} by showing that at each order in the $r$-expansion new undetermined functions appear and further we discuss in section \ref{subsec:commentsansatz} that by making the ansatz for the $r$-expansion more general by adding higher powers of $\log r$ one can construct more solutions. Since one can keep adding arbitrary high powers of $\log r$ at each order in $r$ and since we keep finding new undetermined functions at higher orders in $r$ we observe a breakdown of the FG theorem for AlSch space-times. The underlying features responsible for this are already visible in the simple case of a real scalar field on a Schr\"odinger space-time. We discuss this separately in appendix \ref{app:scalars} which should be read in conjunction with this section \ref{sec:solutions}. We will show in section \ref{subsec:moreconstrained} that by imposing suitable constraints on the solution unique FG type $r$-expansions are recovered.

\subsection{The ansatz}\label{subsec:ansatz}

We will be solving the equations of motion \eqref{eq:eombarg} and \eqref{eq:eomMatbarA} that we repeat here for convenience
\begin{eqnarray}
0=\mbox{Eins}_{\mu\nu} & \equiv & R_{\mu\nu} +2g_{\mu\nu} -\frac{1}{2}F_{\mu\rho}F_\nu{}^\rho+\frac{1}{4}F_{\rho\sigma}F^{\rho\sigma}g_{\mu\nu}-2A_\mu A_\nu\,,\label{eq:Scheom Ein}\\
0=\mbox{Vec}_{\nu} & \equiv & \nabla^\mu F_{\mu\nu} - 4A_\nu\,.\label{eq:Scheom Mat}
\end{eqnarray}
These definitions are not to be confused with \eqref{eq:Einsteineqs-AdSside-A2=0-v2} and \eqref{eq:vectoreq1-AdSside-A2=0-v2}.

Using the radial gauge choice \eqref{eq:radialgauge} and the boundary conditions \eqref{eq:radialgaugeframe fielddecomposition}--\eqref{eq:Ar} we make the following ansatz for the subleading terms\footnote{This ansatz will be further discussed in section \ref{subsec:commentsansatz}. Here we only consider even powers of $r$ and powers of $\log r$ up to order $r^{2n}\log^n r$ where $n=0$ corresponds to the first term in the expansion (there may be an overall power of $r$).}
\begin{eqnarray}
A^a &=& A_{(0)}^a+r^2\log r A^a_{(2,1)}+r^2A^a_{(2)} + O(r^4\log^2r)\label{eq:expansionA} \,,\\
A^r &=& r^3\log r A^r_{(2,1)} + r^3A^r_{(2)} + O(r^5\log^2r)\,,\\
g_{ab} &=& \frac{1}{r^4}g_{(-2)ab}+ \frac{\log r}{r^2}g_{(0,1)ab} + \frac{1}{r^2}g_{(0)ab}+\log^2 r g_{(2,2)ab}+\log r g_{(2,1)ab} \nonumber\\
&&+g_{(2)ab}+O(r^2\log^3 r)\,.\label{eq:expansiong}
\end{eqnarray}
The boundary conditions \eqref{eq:e+}--\eqref{eq:Ar} as well as \eqref{eq:A+}--\eqref{eq:A2} are respected by this ansatz.

The frame fields and tangent space components of the vector field can be expanded as
\begin{eqnarray}
e^{+}_a & = & \frac{1}{r^2}e^{+}_{(0)a} + \log r e^{+}_{(2,1)a}  + e^{+}_{(2)a} + \ldots \,,\label{eq:expansione+}\\
e^{-}_a & = & e^{-}_{(0)a} + \ldots \,,\\
A_{+} & = & 1 + r^2\log r A_{+(2,1)}  + r^2A_{+(2)} + \ldots \,,\\
A_{-} & = & r^2\log r A_{-(2,1)}  + r^2A_{-(2)} + \ldots \,,\label{eq:expansionA-}\\
A_{2} & = & r^2\log r A_{2(2,1)}  + r^2A_{2(2)} + \ldots \,,\label{eq:expansionA2}
\end{eqnarray}
where the dots indicate higher order terms. Since the expansions for the frame fields are not fully determined by the metric (because of local Lorentz transformations) we will not be specific about the structure of these higher orders terms. The expansion for the tangent space components of the vector field $A^a$ can be obtained by computing the expansion for the inverse frame fields.

Further, the boundary conditions and the ansatz for the subleading terms are such that we get
\begin{eqnarray}
A_a & = &A_+ e^+_a+A_- e^-_a= \frac{1}{r^2}A_{(0)a} + O(\log r)\,,\label{eq:A constraint on AdS}\\
g_{\mu\nu}A^\mu A^\nu & = & A_- A_--2A_- A_++A_2 A_2=O(r^2\log r)\,,\label{eq:Asquare constraint on AdS}\\
g_{ab}+e^{+}_a e^{+}_b & = & \frac{1}{r^{2}}\gamma_{(0)ab}+ \dots\,,\label{eq:adsleading term}
\end{eqnarray}
with $A_{(0)a}\equiv\gamma_{(0)ab}A^b_{(0)}=e^+_{(0)a}$ and where $\gamma_{(0)ab}=-e^+_{(0)a}e^-_{(0)b}-e^+_{(0)b}e^-_{(0)a}$ is the non-degenerate AlAdS boundary metric $\gamma_{(0)ab}$. From \eqref{eq:expansiong} and \eqref{eq:expansione+} we obtain
\begin{eqnarray}
g_{(-2)ab} & = & -A_{(0)a}A_{(0)b}\,,\label{eq:g-2}\\
g_{(0,1)ab} & = & -A_{(0)a}e^{+}_{(2,1)b} - A_{(0)b}e^{+}_{(2,1)a}\,,\label{eq:g01}\\
g_{(0)ab} & = & -A_{(0)a}e^{+}_{(2)b} - A_{(0)b}e^{+}_{(2)a}  + \gamma_{(0)ab} \,.\label{eq: relation between g0 and gam0}
\end{eqnarray}
It follows from \eqref{eq: relation between g0 and gam0} and $\gamma_{(0)ab}A^a_{(0)}A^b_{(0)}=0$ that
\begin{equation}
g_{(0)ab}A^a_{(0)}A^b_{(0)}=0\,.\label{eq: A0 null with res g0}
\end{equation}

In appendix \ref{subsec:propsofg0ab} we will show that on-shell we have
\begin{eqnarray}
0 & = & A_{(0)}^ae^+_{(2,1)a}\,,\label{eq:A0e+21}\\
0 & = & A_{(0)}^ae^+_{(2)a}\,.\label{eq:A0e+2}
\end{eqnarray}
From this and \eqref{eq: relation between g0 and gam0} we conclude that
\begin{equation}
\text{det}\,g_{(0)ab}=\text{det}\,\gamma_{(0)ab}\,,
\end{equation}
so that $g_{(0)ab}$ is non-degenerate.

Equation \eqref{eq:expansionA-} requires that we have using \eqref{eq:A0e+21} and \eqref{eq:A0e+2}
\begin{eqnarray}
0 & = & A_{(0)a}A_{(2,1)}^a\,,\label{eq:A0A21}\\
0 & = & A_{(0)a}A_{(2)}^a\,,\label{eq:A0A2}\\
0 & = & A_{(0)a}A^a_{(4,2)}\,.\label{eq:A0A42}
\end{eqnarray}
In order to satisfy \eqref{eq:A constraint on AdS} we need
\begin{equation}\label{eq:A0g22}
A_{(0)}^ag_{(2,2)ab}=0\,.
\end{equation}
Finally, in order to satisfy \eqref{eq:Asquare constraint on AdS} we furthermore need
\begin{eqnarray}
0 & = & A_{(0)}^aA_{(0)}^bg_{(2,1)ab}\,,\label{eq:A0A0g21}\\
0 & = & A_{(0)}^aA_{(0)}^bg_{(2)ab}\,,\label{eq:A0A0g2}\\
0 & = & A_{(0)}^aA_{(0)}^bg_{(4,3)ab}\,,\label{eq:A0A0g43}\\
0 & = & A_{(0)}^aA_{(0)}^bg_{(4,2)ab}\,.\label{eq:A0A0g42}
\end{eqnarray}

\subsection{The radial $TV$ gauge}\label{subsec:radialTVgauge}

Having shown that $g_{(0)ab}$ is non-degenerate it is, from an AlSch perspective, more natural to introduce coordinates such that $g_{(0)ab}$ (rather than $\gamma_{(0)ab}$) is manifestly conformally flat. This is not necessary and we could have equivalently decomposed $g_{(0)ab}$ in a null-bein basis. Like we did in appendix \ref{subsection:thesolutions}. However, for simplicity and for our purposes it will prove convenient to introduce explicit coordinates. We thus partially fix the gauge by writing
\begin{equation}\label{eq: Sch gauge choice H0}
g_{(0)ab}dx^adx^b = 2H_{(0)}dTdV\,,
\end{equation}
with some arbitrary (non-vanishing) function $H_{(0)}=H_{(0)}(T,V)$. It then follows from $A^a_{(0)}$ being null with respect to $g_{(0)ab}$ that $A^T_{(0)}A^V_{(0)}=0$ and we choose
\begin{equation}\label{eq:AT0=0}
A^T_{(0)}=0\,,
\end{equation}
and write from now on
\begin{equation}
A^V_{(0)}=V_{(0)}\,.
\end{equation}
The combination of \eqref{eq: Sch gauge choice H0} and \eqref{eq:AT0=0} is what we shall refer to as the radial $TV$ gauge. Note that we are already working in radial gauge.

It follows from \eqref{eq: relation between g0 and gam0} and \eqref{eq:A0e+2} that $g_{(0)ab}A^b_{(0)}=\gamma_{(0)ab}A^b_{(0)}$ from which we deduce that
\begin{eqnarray}
A_{(0)V} & = & 0\,,\\
A_{(0)T} & = & H_{(0)}V_{(0)}\,.
\end{eqnarray}
Hence using \eqref{eq:g-2} we see that
\begin{equation}
g_{(-2)TT}=-A_{(0)T}A_{(0)T}\,,\qquad g_{(-2)VV}=g_{(-2)TV}=0\,,
\end{equation}
and using \eqref{eq:g01} as well as \eqref{eq:A0A21} we find that
\begin{equation}
g_{(0,1)VV}=g_{(0,1)TV}=0\,.
\end{equation}
Further from \eqref{eq:A0g22} we obtain
\begin{equation}
g_{(2,2)VV}=g_{(2,2)TV}=0 \,.
\end{equation}
and using \eqref{eq:A0A0g21}--\eqref{eq:A0A0g42} we get
\begin{equation}
g_{(2,1)VV}=g_{(2)VV}=g_{(4,3)VV}=g_{(4,2)VV}=0 \,.
\end{equation}
Furthermore from \eqref{eq:A0A21}--\eqref{eq:A0A42} it follows that
\begin{equation}
A_{(2,1)}^T=A_{(2)}^T=A_{(4,2)}^T=0\,.
\end{equation}

Based on the ansatz \eqref{eq:expansionA}--\eqref{eq:expansiong} and the results of this section the ansatz in radial $TV$ gauge becomes up to NLO\footnote{In radial $TV$ gauge we say that an AlSch solution has been expanded up to N$^k$LO when we have expanded $A^T$ up to order $r^{2k+2}$, $A^V$ up to order $r^{2k}$, $A^r$ up to order $r^{2k+1}$, $g_{TT}$ up to order $r^{2k-4}$, $g_{TV}$ up to order $r^{2k-2}$ and $g_{VV}$ up to order $r^{2k}$ as an expansion in $r^2$ (ignoring $\log r$ terms).}
\begin{eqnarray}
A^T & = & r^4\log rA^T_{(4,1)}+r^4A^T_{(4)}+O(r^6\log^3 r)\,,\label{eq:ansatzAT}\\
A^V & = & V_{(0)}+r^2\log r A^V_{(2,1)}+r^2 A^V_{(2)}+O(r^4\log^2 r)\,,\label{eq:ansatzAV}\\
A^r & = & r^3\log r A^r_{(2,1)}+r^3 A^r_{(2)}+O(r^5\log^2 r)\,,\label{eq:ansatzAr}\\
g_{TT} & = & -\frac{1}{r^4}H^2_{(0)}V_{(0)}^2+\frac{\log r}{r^2}g_{(0,1)TT}+O(\log^2 r)\,,\label{eq:ansatzgTT}\\
g_{TV} & = & \frac{1}{r^2}H_{(0)}+\log r g_{(2,1)TV}+g_{(2)TV}+O(r^2\log^3 r)\,,\label{eq:ansatzgTV}\\
g_{VV} & = & r^2\log r g_{(4,1)VV}+r^2g_{(4)VV}+O(r^4\log^4 r)\,.\label{eq:ansatzgVV}
\end{eqnarray}
When we write $O(r^6\log^3 r)$ in \eqref{eq:ansatzAT} and similarly in the other expressions we mean that the next term in the ansatz is at most of order $r^6\log^3 r$ but given this ansatz on-shell tighter bounds may be formulated.

The radial $TV$ gauge in the special case where the AlAdS boundary metric is flat, i.e. $H_{(0)}=1$, can describe both asymptotically Poincar\'e and global Schr\"odinger \cite{Blau:2009gd} coordinates.

\subsection{Properties of the NLO terms}\label{subsec:propssolution}

Starting with the ansatz \eqref{eq:ansatzAT}--\eqref{eq:ansatzgVV} we have shown in appendix \ref{subsection:solvingeom} that up to NLO the expansion reads
\begin{eqnarray}
A^T & = & r^4\log r A^T_{(4,1)}+r^4 A^T_{(4)}+O(r^6\log^3 r)\,,\label{eq:AT}\\
A^V & = & V_{(0)}+r^2 A^V_{(2)}+O(r^4\log^2 r)\,,\label{eq:AV}\\
A^r & = & r^3\log r A_{(2,1)}^r+r^3 A^r_{(2)}+O(r^5\log^2 r)\,,\label{eq:Arup}\\
g_{TT} &=& -\frac{1}{r^4}H^2_{(0)}V_{(0)}^2+ \frac{\log r}{r^2}g_{(0,1)TT} +O(\log^2 r)\,, \label{eq:gTT}\\
g_{TV} &=& \frac{1}{r^2}H_{(0)} +g_{(2)TV}+O(r^2\log^3 r)\,,\label{eq:gTV}\\
g_{VV} &=& O(r^4\log^4 r)\,,\label{eq:gVV}
\end{eqnarray}
with
\begin{eqnarray}
0 & = & \partial_V\left(H_{(0)}V_{(0)}\right)\,,\label{eq:H0A0V}\\
0 & = & \partial_V\left(H_{(0)}^{-1}g_{(2)TV}\right)\,,\label{eq:H0-1g2TV}
\end{eqnarray}
and where the coefficients are constrained by equations \eqref{eq:partialVAr21}--\eqref{eq:VAr21}
and \eqref{eq:partialVAr2}--\eqref{eq:Ar2}. These equations can be written as follows
\begin{eqnarray}
\hspace{-.5cm}g_{(0,1)TT} & = & H_{(0)}V_{(0)}^2\left(-4g_{(2)TV}+\partial_T\partial_V\log H_{(0)}\right)\,,\label{eq: second order coeff1}\\
\hspace{-.5cm}A_{(2)}^V & = & \frac{V_{(0)}}{H_{(0)}}\left(-g_{(2)TV}+\frac{1}{4}H_{(0)}\left(V_{(0)}\partial_V\left(V_{(0)}\partial_V\right)-8\right)U_{(4)}\right)\,,\\
\hspace{-.5cm}A^T_{(4,1)} & = & -\frac{1}{2H_{(0)}V_{(0)}}\left(V_{(0)}\partial_V\left(V_{(0)}\partial_V\right)-8\right)U_{(4)}\,,\label{eq:AT41}\\
\hspace{-.5cm}A_{(4)}^T & = & \frac{1}{4H^2_{(0)}V_{(0)}}\left(2g_{(2)TV}-\partial_T\partial_V\log H_{(0)}+4H_{(0)}U_{(4)}\right)\,,\label{eq:defU4}\\
\hspace{-.5cm}A^r_{(2,1)} & = & -\frac{1}{4}V_{(0)}\partial_V\left(\left(V_{(0)}\partial_V\left(V_{(0)}\partial_V\right)-8\right)U_{(4)}\right)\,,\\
\hspace{-.5cm}A^r_{(2)} & = & \frac{1}{8}V_{(0)}\partial_V\left(4U_{(4)}-\left(V_{(0)}\partial_V\left(V_{(0)}\partial_V\right)-8\right)U_{(4)}+\frac{1}{H_{(0)}}\partial_T\partial_V\log H_{(0)}\right)\,,\label{eq: second order coeff6}
\end{eqnarray}
where we introduced a new function $U_{(4)}$ defined by relation \eqref{eq:defU4}. These relations express the coefficients in terms of $V_{(0)}$, $g_{(2)TV}$, $H_{(0)}$ and $U_{(4)}$. The latter two are by equations \eqref{eq:partialVAr21}, \eqref{eq:EinsTrorder r-1} and \eqref{eq:partialVAr2} constrained to satisfy
\begin{eqnarray}
0 & = & \left(V_{(0)}\partial_V\left(V_{(0)}\partial_V\right)-8\right)^2U_{(4)}\,,\label{eq:U4}\\
0 & = & \partial_V\left(V_{(0)}\partial_V\left(V_{(0)}\partial_T\partial_V\log H_{(0)}\right)\right)\label{eq: constraint H0}\,.
\end{eqnarray}
It turns out that the metric coefficients in $g_{TV}$ at order $r^2$ (including the log terms) and in $g_{VV}$ at order $r^4$ (including the log terms) are also fully determined as functions of $V_{(0)}$, $g_{(2)TV}$, $H_{(0)}$ and $U_{(4)}$.

The metric expansions \eqref{eq:gTT}, \eqref{eq:gTV} and \eqref{eq:gVV} can be written in terms of our frame field basis \eqref{eq:radialgaugeframe fielddecomposition} as follows
\begin{eqnarray}
e^{+}_{T} &=& \frac{1}{r^2}H_{(0)}V_{(0)}+ \log r e^+_{(2,1)T}+ \dots \,,\\
e^{-}_{V} &=& -\frac{1}{V_{(0)}}+r^2\log r e^-_{(2,1)V}+r^2 e^-_{(2)V}+ \dots \,,
\end{eqnarray}
where we used \eqref{eq:adsleading term} to derive the form of $e^{-}_{T}$ and where
\begin{eqnarray}
e^+_{(2,1)T} & = & -\frac{1}{2}\frac{g_{(0,1)TT}}{H_{(0)}V_{(0)}}\,,\\
e^-_{(2,1)V} & = & -\frac{1}{2}\frac{g_{(0,1)TT}}{H^2_{(0)}V_{(0)}^3}\,,\\
e^-_{(2)V} & = & -\frac{g_{(2)TV}}{H_{(0)}V_{(0)}}\,.
\end{eqnarray}
The components $e^{-}_{T}$ and $e^{+}_{V}$ are not needed at the order at which we are working. They are of orders $r^2$ and $r^4$, respectively up to logarithmic corrections.

Comparing the expansion for $e^+_T$ with \eqref{eq:expansione+} we find that $e^+_{(2)a}=0$ so that using \eqref{eq: relation between g0 and gam0} we conclude that\footnote{We mention that there are other choices possible for the subleading terms of the frame field (involving $e^-_T$) such that we reproduce the same metric expansion but such that $g_{(0)ab}$ and $\gamma_{(0)ab}$ differ in their $TT$ component.}
\begin{equation}\label{eq:g0=gamma0}
g_{(0)ab}=\gamma_{(0)ab}\,.
\end{equation}
Hence the metric coefficient $g_{(0)ab}$ appearing at order $r^{-2}$ in the expansion of $g_{ab}$ is equal to the boundary metric of the associated AlAdS metric $\gamma_{ab}=g_{ab}+e^+_a e^+_b$. In the following whenever we speak of the boundary metric we will be referring to the metric $\gamma_{(0)ab}$ as well as we will have in mind a choice of frame fields that is such that \eqref{eq:g0=gamma0} holds.

The AlAdS metric $\gamma_{ab}$ has the following expansion
\begin{equation}
\gamma_{TV} = \frac{1}{r^2}H_{(0)}+g_{(2)TV}+\dots\,,\\
\end{equation}
and where $\gamma_{TT}$ is of order $r^0$ and $\gamma_{VV}$ of order $r^4$ up to logarithmic corrections. To solve for the coefficient in the expansion of $\gamma_{TT}$ at order $r^0$ is already quite a daunting task as the equations to be solved for become quite intricate at the next order in the expansion of the equations of motion discussed in section \ref{subsection:solvingeom}.

\subsection{Comments}\label{subsec:higher}\label{subsec:commentsansatz}

If we were to expand to higher orders, i.e. solve for the higher order coefficients in the ansatz \eqref{eq:ansatzAT}--\eqref{eq:ansatzgVV}, then we would find that the coefficients have to obey relations of the type \eqref{eq:U4} involving $\partial_V$ derivatives. The solutions would lead to free integration constants that are arbitrary functions of $T$. Since this happens at each order we have an infinite number of free functions of $T$ that appear in the expansion and that are not fixed by the equations of motion. This is not in the spirit of a FG expansion. A similar behavior can be observed in the case of a scalar field on a fixed Schr\"odinger background. We discuss this case in appendix \ref{app:scalars}.

Apart from the issue that we have to specify infinitely many functions (and not just a source and a vev as is the case for AlAdS backgrounds) we also cannot write down the most general $r$-expansion that covers all solutions. So all solutions will depend on the ansatz that has been made. This phenomenon also has a counterpart for the case of a scalar field on a fixed Schr\"odinger background as discussed in appendix \ref{app:scalars}.

We conclude that in full generality the radial dependence is not fixed. In fact we find that the $r$-dependence of the expansion and the $V$-dependence of the coefficients are correlated. By this we mean that by allowing for more logs in the $r$-expansion we can allow for more general $V$-dependences of the coefficients. This can already be seen within the ansatz \eqref{eq:ansatzAT}--\eqref{eq:ansatzgVV}. Since setting to zero e.g. $A_{(4,1)}^T$ implies
\begin{equation}
\left(V_{(0)}\partial_V\left(V_{(0)}\partial_V\right)-8\right)U_{(4)}=0\,,
\end{equation}
as follows from \eqref{eq:AT41}. This is to be contrasted with equation \eqref{eq:U4} whose solution has a more general $V$ dependence.
This supports the claim that the powers of the log terms in the ansatz are related to the $V$-dependence of the solution and the higher the power of the log terms the more freedom we find for the $V$-dependence of our solution.

We thus conclude that we cannot write the most general $r$-expansion that covers all solutions as a power series in $r$ because at each order in $r$ we can add arbitrary high powers of $\log r$ and that furthermore within a given ansatz new undetermined functions keep appearing at higher orders. We thus conclude that one cannot in general write down a FG expansion that covers the most general asymptotic solution to the equations of motion \eqref{eq:Scheom Ein} and \eqref{eq:Scheom Mat} of the massive vector model with AlSch boundary conditions.

In the following sections we will show that there are ways around this problem by imposing suitably chosen constraints. We will say that a radial expansion is of FG type whenever the problems encountered in this and the previous subsection do not arise. We will not test the robustness of the expansions against more general ans\"atze than those that can be written as a series in $r^{2n}\log^m r$ with $m$ and $n$ non-negative integers.

\section{Fefferman--Graham expansions for constrained solutions}\label{subsec:moreconstrained}

Inspired by the computation done in section \ref{subsec:constrained} where we imposed the constraint $g_{\mu\nu}A^\mu A^\nu=0$ and obtained FG type solutions we introduce in this subsection new and weaker constraints that bypass the problems mentioned in the previous section related to the dependence on the ansatz
and the appearance of an infinite amount of free functions. We will first recover
the $g_{\mu\nu}A^\mu A^\nu=0$ solution from an AlSch perspective and then consider two other
ways to constrain the solutions by:
\begin{enumerate}
\item demanding the full solution to be $V$-independent,
\item demanding that $\partial_V$ is a null vector everywhere.
\end{enumerate}
In both cases we obtain FG type solutions
that are robust against making more general ans\"atze. By
this we mean that the equations of motion supplemented with the respective
constraints will no longer allow for arbitrary high powers of $\log r$. We can then
check explicitly that nothing can be found beyond our initial ansatz and the latter
will cease to be a critical assumption. Further, the asymptotic expansions have a finite number of free functions.

\subsection{$A^\mu$ is null: an AlSch point of view}\label{subsubsec:A^2=0again}
As expected by solving the equations of motion with an ansatz of the form \eqref{eq:ansatzAT}--\eqref{eq:ansatzgVV}
using the constraint $g_{\mu\nu}A^\mu A^\nu=0$ we reproduce
exactly the result, equations \eqref{eq:solAdSsideg2}--\eqref{eq:solAdSsideAr2}, obtained previously from an AlAdS perspective. In radial $TV$ gauge the (exact) solution reads
\begin{eqnarray}
A^V &=& V_{(0)} + r^4A^V_{(4)}\,,\\
g_{TT}  &=& -\frac{1}{r^4}V_{(0)}^2 + g_{(2)TT} -\frac{1}{3}r^4(A^V_{(4)})^2\,, \label{eq:gTTA^2=0again}\\
g_{TV}  &=& \frac{1}{r^2}\,,\\
g_{VV}  &=& A^T = A^r = 0\,,
\end{eqnarray}
where the free functions are arbitrary functions of the time coordinate only, i.e.
$V_{(0)}=V_{(0)}(T)$, $A^V_{(4)}=A^V_{(4)}(T)$ and
$g_{(2)TT}=g_{(2)TT}(T)$. In this case we find that $H_{(0)}$ factorizes and we have chosen to set it equal to one. We conclude that the solution with $A^\mu$ null terminates at N$^4$LO.

These solutions with $A_\mu A^\mu=0$ also have the property that
\begin{equation}
F_{\mu\nu}F^{\mu\nu}=0\,.
\end{equation}
They are therefore also solutions of any theory containing scalar fields on top of the massive vector field with the scalars set equal to constant values. This is true even if the scalars couple to $F_{\mu\nu}F^{\mu\nu}$ and $A_\mu A^\mu$. This is because the scalar equations of motion are solved by $A_\mu A^\mu=F_{\mu\nu}F^{\mu\nu}=0$ when we take the scalars constant. In particular they are solutions of the 3-dimensional supergravity action given in
\cite{Detournay:2012dz}.

The $A_\mu A^\mu=0$ solutions with $V_{(0)}$ constant have been constructed in \cite{Nakayama:2010xq}. It would be interesting to study the full class of $A_\mu A^\mu=0$ solutions in more detail as they are exact and generic solutions that can possibly describe interesting time dependent phenomena. Relatedly it would be interesting to know the full class of solutions with $A^\mu$ null in arbitrary dimensions\footnote{There is a straightforward generalization of the 3-dimensional exact solutions with $A^\mu$ null to $d+3$ dimensions where the action is given by (possibly dressed with scalar fields)
\begin{equation}
S=\int d^{d+3}x\sqrt{-g}\left(R-\frac{1}{4}F^2-(d+2)A^2+(d+1)(d+2)\right)\,.
\end{equation}
The exact solution in $d+3$ dimensions is (see also \cite{Nakayama:2010xq})
\begin{eqnarray}
g_{TT} &=& -\frac{V^2_{(0)}}{r^4} +r^d g_{(d)TT}-\frac{1}{2}\frac{d+2}{d+3}r^{2d+4}V^2_{(d+4)} \,,\\
g_{TV} &=& \frac{1}{r^2}\,,\\
A^V &=& V_{(0)} + r^{d+4}V_{(d+4)}\,,
\end{eqnarray}
with all other components zero and where the coefficients are functions of $T$. We are not claiming that in higher dimensions these are the only solutions satisfying the constraint that $A^\mu$ is null.}.

The solutions of this section have the property that the generator of particle number $N=\partial_V$ is a null Killing vector. We will next discuss two generalizations in which 1). $N$ is a Killing vector but not necessarily null and 2). in which $N$ is null but not necessarily Killing.

\subsection{$\partial_V$ is a Killing vector}\label{subsubsec:Vindependentsolutions}

In order to enlarge the solution space we are going to relax some of the properties of the solution of the previous section. The property that we will not enforce here is that $A^\mu$ must be null. Instead we demand that in the radial $TV$ gauge $\partial_V$ is a Killing vector.

As it turns out there are two classes of solutions that have $\partial_V$ as a Killing vector. The first class is the solution with $A^\mu$ null and $\partial_T V_{(0)}\neq 0$. The second class must have $\partial_T V_{(0)}=0$ and coincides with the $A^\mu$ is null solution at low orders in the expansion. To make it clear that there are two such cases we start by demanding a slightly less restrictive property namely that the full solution be linear in $V$.  We will show that such solutions are necessarily independent of the $V$ coordinate with the two possibilities as just described.

The constraint that the metric $g_{\mu\nu}$ and the vector field $A^\mu$ are at most linear in $V$ forces us to take $H_{(0)}$ to be $V$-independent so that we can again take it to be unity. Solving the equations of motion up to NNLO in $r$ we find that the most general solution linear in $V$ is given by
\begin{eqnarray}
\hspace{-.7cm}A^T &=& r^4A^T_{(4)}+r^6A^T_{(6)} +\dots \,,\\
\hspace{-.7cm}A^V &=& V_{(0)} + r^2A^V_{(2)} + r^4\log r A^V_{(4,1)}+r^4 A^V_{(4)} + \dots \,,\\
\hspace{-.7cm}A^r &=& r^5A^r_{(4)} + \dots\,,\\
\hspace{-.7cm}g_{TT} &=& -\frac{1}{r^4}V_{(0)}^2+\frac{\log r}{r^2}g_{(0,1)TT}+\log r g_{(2,1)TT}+g_{(2)TT} +\dots \,,\\
\hspace{-.7cm}g_{TV} &=& \frac{1}{r^2} + g_{(2)TV}+r^2g_{(4)TV}+ \dots\,,\\
\hspace{-.7cm}g_{VV} &=& r^4g_{(6)VV} + \dots\,,
\end{eqnarray}
where
\begin{eqnarray}
A^T_{(4)} & = & \frac{1}{2}\frac{g_{(2)TV}}{V_{(0)}}\,,\\
A^V_{(2)} & = & - V_{(0)}g_{(2)TV}\,,\\
A^V_{(4,1)} & = & V_{(0)}\left(g_{(2)TV}\right)^2\,,\\
A^r_{(4)} & = & -\frac{1}{2}\partial_VA^V_{(4)}-\frac{1}{4}\partial_T\left(\frac{g_{(2)TV}}{V_{(0)}}\right)\,,\\
g_{(0,1)TT} & = & -4V_{(0)}^2g_{(2)TV}\,,\\
g_{(2,1)TT} & = & -6V_{(0)}^2\left(g_{(2)TV}\right)^2\,.
\end{eqnarray}
The free functions are $V_{(0)}(T)$, $g_{(2)TV}(T)$, $g_{(2)TT}$ and $A^V_{(4)}$ where the latter two satisfy
\begin{eqnarray}
\partial_V g_{(2)TT} &=& \frac{g_{(2)TV}}{V_{(0)}}\partial_TV_{(0)}\,,\label{eq:h2TT}\\
\partial_VA^V_{(4)} & = & A^V_{\text{lin}(4)}(T)\,.\label{eq:AV4}
\end{eqnarray}
The coefficients $A^T_{(6)}$, $g_{(4)TV}$ and $g_{(6)VV}$ are functions of $g_{(2)TV}$ and $V_{(0)}$ but we do not give their explicit form.

This solution is well behaved in the sense that making the ansatz more general by adding higher powers of $\log r$ does not lead to new solutions and there are no additional free functions on top of the functions $V_{(0)}$, $g_{(2)TV}$, $g_{(2)TT}$ and $A^V_{(4)}$ appearing at higher orders. Here this is achieved by forcing the solution to be linear in $V$ to all orders. However, by solving at higher and higher orders some constraints do appear on the lower orders. In particular at N$^4$LO we find that there is a coefficient that is
proportional to $(A^V_{\text{lin}(4)})^2V^2$. By requiring that the solution must remain linear in $V$ we obtain the condition $A^V_{\text{lin}(4)}=0$. At N$^6$LO we find that there is a coefficient that is proportional to $(g_{(2)TV})^4(\partial_TV_{(0)})^2V^2$ so that imposing a linear $V$-dependence constraints the solution further. Hence by expanding the fields up to N$^6$LO and by solving the equations of motion we have obtained the following two constraints
\begin{equation}
(A^V_{\text{lin}(4)})^2=(g_{(2)TV})^4(\partial_TV_{(0)})^2 = 0\,.
\end{equation}
Enforcing these constraints kills all the terms linear in $V$ and leaves us with a $V$-independent solution.

We thus conclude that the most general $V$-independent solution splits in two distinct cases:
\begin{enumerate}
\item For $g_{(2)TV}=A^V_{\text{lin}(4)}=0$ the solution can be seen to obey the $g_{\mu\nu}A^\mu A^\nu=0$ constraint
and is hence exactly the same as the one we found in \ref{subsubsec:A^2=0again}.
\item For $\partial_TV_{(0)}=A^V_{\text{lin}(4)}=0$ we can take $V_{(0)}=1$ keeping $H_{(0)}=1$. This is then the most general solution independent of $V$ with $g_{\mu\nu}A^\mu A^\nu\neq0$.
\end{enumerate}
The solution with $g_{(2)TV}\neq0$ up to NNLO is
\begin{eqnarray}
A^T &=& r^4A^T_{(4)}+r^6A^T_{(6)}+ \dots \,,\\
A^V &=& 1+r^2A^V_{(2)}+r^4\log r A^V_{(4,1)} + r^4A^V_{(4)} + \dots \,,\\
A^r &=& r^5A^r_{(4)}+ \dots\,,\\
g_{TT} &=& -\frac{1}{r^4}+\frac{\log r}{r^2}g_{(0,1)TT}+\log r g_{(2,1)TT}+g_{(2)TT}+ \dots \,,\\
g_{TV} &=& \frac{1}{r^2} + g_{(2)TV} + r^2g_{(4)TV}  + \dots\,,\\
g_{VV} &=& r^4g_{(6)VV}+ \dots\,,
\end{eqnarray}
where
\begin{eqnarray}
A^T_{(4)} & = & \frac{1}{2}g_{(2)TV}\,,\\
A^V_{(2)} & = & -g_{(2)TV}\,,\\
A^r_{(4)} & = & -\frac{1}{4}\partial_Tg_{(2)TV}\,,\\
g_{(0,1)TT} & = & -4g_{(2)TV}\,.
\end{eqnarray}
The coefficients $A^T_{(6)}$, $A^V_{(4,1)} $, $g_{(2,1)TT}$, $g_{(4)TV}$ and $g_{(6)VV}$ are all proportional to $(g_{(2)TV})^2$. The free functions are $g_{(2)TV}(T)$ and $V_{(4)}(T)$. Upon setting $g_{(2)TV}=0$ we recover the $g_{\mu\nu}A^\mu A^\nu=0$ solution with $V_{(0)}=1$. In order to see also the term $ -\frac{1}{3}r^4(A^V_{(4)})^2$ in the expansion of $g_{TT}$ for the case that $A^\mu$ is null requires that we expand the above solution up to N$^4$LO and then set $g_{(2)TV}=0$.

This class of solutions is very reminiscent of the class of AlSch solutions obtained via TsT in a supergravity context. TsT transformations resulting in an AlSch metric by starting with an AlAdS metric can only be done when the particle number generator is a Killing vector.

\subsection{$\partial_V$ is null}\label{sec: A^T=0 sol}

The solutions presented in the previous section are of the standard FG type. There is a unique radial expansion containing a finite number of undetermined functions. So far in the literature such expansions have only been written down whenever there was a Killing vector present. We will now construct a class of solutions that are of the FG type without any Killing vectors present. This class is obtained by demanding that $\partial_V$ is null so that $g_{VV}=0$. The easiest way to obtain this class of solutions is to start with the class presented in \ref{subsec:propssolution} and to enforce the condition that $A^T=0$. This will eventually lead to solutions that all have $g_{VV}=0$ everywhere. We then check that on-shell demanding  $A^T=0$ is equivalent to putting $g_{VV}=0$.

As can be seen  from the NLO solution \eqref{eq:AT}--\eqref{eq:gVV} the condition $A^T_{(4,1)}=0$ implies
\begin{equation}
\left(V_{(0)}\partial_V\left(V_{(0)}\partial_V\right)-8\right)U_{(4)}=0\,,\label{eq: weaker constraint on H0}
\end{equation}
so that
\begin{equation}
A^V_{(2)}=-\frac{V_{(0)}g_{(2)TV}}{H_{(0)}}\,.
\end{equation}
Setting $A^T_{(4)}=0$ in \eqref{eq:defU4} together with equation \eqref{eq: constraint H0} for $H_{(0)}$ as well as \eqref{eq:H0A0V} and \eqref{eq:H0-1g2TV} gives
\begin{equation}
\partial_V\left(V_{(0)}\partial_VU_{(4)}\right) = 0\,,
\end{equation}
which combined with \eqref{eq: weaker constraint on H0} implies
\begin{equation}
U_{(4)}=0\,.
\end{equation}
Plugging this into \eqref{eq:defU4} and using $A^T_{(4)}=0$ we obtain
\begin{equation}\label{eq: constraint H0 for AT=0}
\partial_T\partial_V\log H_{(0)} = 2g_{(2)TV}\,.
\end{equation}
Solving the equations of motion at higher order keeping $A^T=0$ we find a unique expansion which up to NNLO reads
\begin{eqnarray}
A^V &=& V_{(0)} + r^2A^V_{(2)} + r^4A^V_{(4)}  + \ldots\,, \\
A^r &=& r^5A^r_{(4)}  + \ldots\,,\\
g_{TT}  &=& -\frac{1}{r^4}H^2_{(0)}V_{(0)}^2 + \frac{\log r}{r^2}g_{(0,1)TT} +\log r g_{(2,1)TT} +g_{(2)TT} + \ldots \,,\\
g_{TV}  &=& \frac{1}{r^2}H_{(0)}+g_{(2)TV}+r^2g_{(4)TV} +\ldots\,,\\
g_{VV}  &=& 0\,,
\end{eqnarray}
with
\begin{eqnarray}
A^V_{(2)}&=& -\frac{V_{(0)}g_{(2)TV}}{H_{(0)}}\,,\\
A^r_{(4)} & = & -\frac{1}{2H_{(0)}}\partial_V\left(H_{(0)}A^V_{(4)}\right)\,,\\
g_{(0,1)TT} & = & -2 H_{(0)}V_{(0)}^2 g_{(2)TV}\,,\\
g_{(2,1)TT} & = & -2\left(V_{(0)}g_{(2)TV}\right)^2\,,\\
g_{(4)TV} & = & \frac{\left(g_{(2)TV}\right)^2}{4H_{(0)}}\,.
\end{eqnarray}
The free functions that determine the full solution (also at higher orders) are $H_{(0)}$, $V_{(0)}$, $g_{(2)TV}$, $g_{(2)TT}$ and $A^V_{(4)}$ subject to \eqref{eq: constraint H0 for AT=0}, \eqref{eq:H0A0V}, \eqref{eq:H0-1g2TV} and
\begin{eqnarray}
\partial_V g_{(2)TT} &=& H_{(0)}\partial_T\left(\frac{g_{(2)TV}}{H_{(0)}}\right)\,,\\
0 &=& \partial_V\left(\frac{1}{H_{(0)}}\partial_V\left(H_{(0)}A^V_{(4)}\right)\right)\,.
\end{eqnarray}

These solutions depend explicitly on $V$ in two different ways. First of all through $H_{(0)}$ so that for $\partial_VH_{(0)}\neq 0$ particle number is already broken on the boundary, but even when we take a flat boundary so that $H_{(0)}=1$ we will find in the expansion terms depending on $A^V_{(4)}$ (and powers thereof) with $A^V_{(4)}$ linear in $V$. The latter case has $\partial_V$ as an asymptotic null Killing vector and so particle number is a symmetry near the boundary. This is the first example of a class of AlSch solutions that breaks particle number in the IR.

\section{Holographic renormalization}\label{sec:HR}

For all AlAdS$_3$ space-times of AdS$_3$ gravity without matter fields the following bulk plus Gibbons--Hawking plus counterterm action suffices for the purpose of holographic renormalization \cite{Balasubramanian:1999re,deHaro:2000xn}
\begin{equation}\label{eq:AdScounterterms}
S=\int d^3x\sqrt{-g}\left(R+2\right)+2\int
d^2\xi\sqrt{-h}\left(K-1+\frac{1}{2}R_{(h)}\log r\right)\,,
\end{equation}
where we use radial gauge with the boundary at $r=0$ and where $R_{(h)}$ is the Ricci scalar of the metric $h_{ab}=g_{ab}$ and where
\begin{equation}
h=\text{det}\,g_{ab}\,.
\end{equation}
In this case we have a FG expansion for all solutions with the boundary metric fully arbitrary. This enables us to compute the on-shell variation with respect to varying
the boundary metric. In the case of AlSch space-times we do not have full control over the complete set of solutions to the equations of motion of the massive vector
model with AlSch boundary conditions so we cannot expect to compute something like the on-shell variation with respect to a certain source. However we can make a view simple
observations about the behavior of the on-shell action for the solutions discussed in the previous section.

Starting with the solutions described in section \ref{subsec:moreconstrained} that have a well behaved FG expansion we notice that they all have $U_{(4)}=0$. We recall that $U_{(4)}$ has been defined in \eqref{eq:defU4}. For all the solutions of section \ref{subsec:moreconstrained} we find that
\begin{equation}
S = \int d^3x \sqrt{-g}\left( R +2 - \frac{1}{4}F_{\mu\nu}F^{\mu\nu} -2A_\mu A^\mu \right) +2\int
d^2\xi\sqrt{-h}\left(K-1+\frac{1}{2}R_{(h)}\log r\right)\,,\label{eq: HR action U4 zero}
\end{equation}
is finite on-shell where we used only the AdS$_3$ counterterms\footnote{Also in the case of the class of 5-dimensional AlSch space-times that can be obtained via TsT it suffices to just add the usual AlAdS$_5$ counterterms \cite{Hartong:2010ec}.}. Although for the NLO solution \eqref{eq:AT}--\eqref{eq:gVV} both $F_{\mu\nu}F^{\mu\nu}$ and $A_\mu A^\mu$ do contribute divergently to the on-shell action it can be seen that the combination $-\frac{1}{4}F_{\mu\nu}F^{\mu\nu} -2A_\mu A^\mu$ does not if $U_{(4)}=0$. We note
that for the purposes of computing the divergences of the on-shell action for the class of solutions defined by our ansatz \eqref{eq:ansatzAT}--\eqref{eq:ansatzgVV} it is sufficient to know the solution up to NLO. In this section whenever we talk about the on-shell value of some quantity we will always mean so within the context of the ansatz \eqref{eq:ansatzAT}--\eqref{eq:ansatzgVV}.

If however we evaluate the on-shell action \eqref{eq: HR action U4 zero} for $U_{(4)}\neq 0$, i.e. for the full NLO solution \eqref{eq:AT}--\eqref{eq:gVV} the result is divergent. More precisely we find that the action contains a
$\log^2 r$ and a $\log r$ divergence which are given by
\begin{eqnarray}
\left.S\right|_{{\tiny\mbox{on-shell}}} &=& \log^2 r \int dTdV  \left(
-2H_{(0)}^2V_{(0)}^2\partial_V\left(\frac{\partial_V
U_{(4)}}{H_{(0)}}\right)+16H_{(0)}U_{(4)}\right) \nonumber\\
&&+\log r \int dTdV 8 H_{(0)}U_{(4)}+O(1) \,.\label{eq: onshell action div}
\end{eqnarray}

Before proceeding we make a few remarks about this result. First of all, it can be noted that using \eqref{eq:U4} the terms $H_{(0)}U_{(4)}$ can be rewritten as the $V$-derivative of something, so we can think of it as the divergence of something. Secondly, the presence of the $\log^2 r$ divergence is of course due to our choice of ansatz, but since we can make more general ans\"atze involving higher powers of $\log r$ as shown in section \ref{subsec:commentsansatz} we expect that arbitrary high powers of $\log r$ can appear in \eqref{eq: HR action U4 zero} by including more and more logs in the ansatz.

We now consider the addition of extra counterterms in order to try to remove the divergences. We will list all such terms and organize them according to the number $m$ of derivatives and number $n$ of vectors $A^a$ they have. We indicate these numbers in the counterterm $\mathcal{T}_{[m,n]}$ as subscripts with square brackets to avoid confusion with the indication of the orders of the fields. Because on the boundary $\sqrt{-h}$ is of order $r^{-2}$ it follows that any term which goes to zero faster than $r^2$ will not contribute. Counterterms that have an odd number of derivatives also have an odd number of derivatives on-shell and can therefore not be used to remove the divergences in \eqref{eq: onshell action div} which is even in derivatives as there is no mechanism by which an expression can change from an odd to an even number of
derivatives.

There is only one term that is second order in derivatives and that does not contain any vector fields which is $\sqrt{-h}R_{(h)}$. Its on-shell value is
\begin{equation}
\mathcal{T}_{[2,0]} = \sqrt{-h}R_{(h)} = -2\partial_T\partial_V\log H_{(0)} +O(r^2\log^2r)\,.\label{eq: CT Rh0}
\end{equation}
Then there is one term that is zeroth order in derivatives and second order in the vector field. This term is $h_{ab}A^a A^b$ and on-shell we have
\begin{eqnarray}
\mathcal{T}_{[0,2]} = \sqrt{-h}h_{ab}A^aA^b &=& \log r \left(
-H_{(0)}^2V_{(0)}^2\partial_V\left(\frac{\partial_V U_{(4)}}{H_{(0)}}\right)
+8H_{(0)}U_{(4)} \right) \nonumber\\ &&
+g_{(2)TV}-\frac{1}{2}\partial_T\partial_V\log H_{(0)}
+ 2H_{(0)}U_{(4)} + O(r^2\log^4r)\,.\label{eq: CT A^2}
\end{eqnarray}

At second order in derivatives and second order in the boundary vector field we will first form all possible divergences\footnote{We note that we are really performing the holographic renormalization at the level of the boundary Lagrangian.}  of the form
$\sqrt{-h}\nabla^{(h)}_a\mathcal{J}_{[1,2]}^a$ where $\mathcal{J}_{[1,2]}^a$ is any linear combination of
\begin{equation}
A^a \nabla^{(h)}_bA^b\,,\quad \nabla^{(h)a}A^2_{(h)}\,,\quad X^a_{(h)} \,,\label{eq: J12}
\end{equation}
where as before $X^a_{(h)}=A^b\nabla^{(h)}_bA^a$ with $\nabla^{(h)}$ the $h_{ab}$ covariant derivative and where we defined
\begin{equation}
A^2_{(h)}=h_{ab}A^aA^b\,.
\end{equation}
Later we will consider all other possibilities modulo these total derivatives. The vector $\mathcal{J}_{[1,2]}^a$ is the most general vector consisting of two vector fields and one derivative. On-shell it can be seen that the first and the last term of the set \eqref{eq: J12} give the same contribution. We will keep only the last one. Specifically we have
\begin{eqnarray}
\sqrt{-h}\square^{(h)}A^2_{(h)} &=& 8\log r \left(
-H^2_{(0)}V^2_{(0)}\partial_V\left(\frac{\partial_V U_{(4)}}{H_{(0)}}\right)
+8H_{(0)}U_{(4)} \right) \nonumber\\ && +
2H^2_{(0)}V^2_{(0)}\partial_V\left(\frac{\partial_V U_{(4)}}{H_{(0)}}\right) +
O(r^2\log^4r)\,,\label{eq: contribution BoxA^2}\\
\sqrt{-h}\nabla_a^{(h)}X^a_{(h)} &=& 2H^2_{(0)}V^2_{(0)}\partial_V\left(\frac{\partial_V
U_{(4)}}{H_{(0)}}\right) - 16H_{(0)}U_{(4)}+
O(r^2\log^4r)\,.\label{eq: contribution
DivXh}
\end{eqnarray}
All possible counterterms of the type $\mathcal{T}_{[2,2]}$ can then be constructed
by taking linear combinations of elements in the set
\begin{equation}
\mathcal{T}_{[2,2]} \in \sqrt{-h}\left\{ \nabla^{(h)}_a A_b \nabla^{(h)a}A^b
\,,\; \nabla^{(h)}_a A_b \nabla^{(h)b}A^a\,,\; (\nabla^{(h)}_a A^a)^2\,,\;
\nabla^{(h)}_a\mathcal{J}_{[1,2]}^a\right\}  \,.\label{eq: C22}
\end{equation}
Evaluating the first three terms in \eqref{eq: C22} it can be seen that none of them
contribute to the on-shell action. Hence we conclude that \eqref{eq: contribution BoxA^2} and \eqref{eq:
contribution DivXh} are the only relevant terms in $\mathcal{T}_{[2,2]}$.

There are various ways of removing the $\log^2 r$ divergence using the counterterms discussed above. Two candidates are $\sqrt{-h}h_{ab}A^aA^b$ and $\sqrt{-h}\square^{(h)}A^2_{(h)}$ both multiplied by $\log r$. We choose the second option and use $\sqrt{-h}\square^{(h)}A^2_{(h)}$ again but not multiplied by $\log r$ to simplify the remaining $\log r$ divergence as much as possible. Specifically we take
\begin{equation}
S = S_{{\tiny\mbox{bulk}}} +2\int
d^2\xi\sqrt{-h}\left(K-1-\frac{1}{32}\square^{(h)}A^2_{(h)}+\log
r\left(\frac{1}{2}R_{(h)}-\frac{1}{8}\square^{(h)}A^2_{(h)}\right)\right)\,,
\end{equation}
whose on-shell value gives
\begin{eqnarray}\label{eq: on-shell action div removed1}
\left.S\right|_{{\tiny\mbox{on-shell}}} &=& 4\log r \int dTdV
H_{(0)}U_{(4)} + O(1)\,.
\end{eqnarray}
We are missing a counterterm whose independent contribution is proportional to $H_{(0)}U_{(4)}$.

We next consider terms with two derivatives but fourth order in the vector field. Starting with terms that can be written as a divergence we look at terms of the form
$\sqrt{-h}\nabla^{(h)}_a\mathcal{J}_{[1,4]}^a$ where $\mathcal{J}_{[1,4]}^a$ consists of a linear combination of all possible contractions of terms of the form
\begin{equation}
A A A\nabla^{(h)}A\,.\label{eq: J14}
\end{equation}
There are 4 inequivalent index contractions. It turns out that only one of these, namely $A^aA^b A^c\nabla^{(h)}_bA_c$ can contribute. However its contribution is equal to that of \eqref{eq: contribution BoxA^2} already found previously. The remaining terms in $\mathcal{T}_{[2,4]}$ that are not a divergence can be constructed from all possible contractions of
\begin{equation}
\sqrt{-h} A A\nabla^{(h)}A\nabla^{(h)}A \,.
\end{equation}
It can be seen that all of these are at least $O(r^2\log^4 r)$ and hence again cannot be used to remove any
divergence of the on-shell action.

Since upon making use of the relation \eqref{eq:U4} the term in \eqref{eq: on-shell action div removed1} is
equal to
\begin{equation}
H_{(0)}U_{(4)} = H_{(0)}V_{(0)}\partial_V\left[ -
\frac{1}{64}V_{(0)}\partial_V\left(V_{(0)}\partial_V\left(V_{(0)}\partial_VU_{(4)}\right)\right)
+ \frac{1}{4}V_{(0)}\partial_VU_{(4)}\right]\,.\label{eq: diff rel H0U4}
\end{equation}
we may try to find additional counterterms that are fourth order in derivatives.

We will now list all the terms that can contribute at fourth order in derivatives. The terms in $\mathcal{T}_{[4,0]}$ consist of $\sqrt{-h}\square^{(h)} R_{(h)}$ and
$\sqrt{-h}R^2_{(h)}$ both of which do not contribute. As before, we first construct all divergences
$\sqrt{-h}\nabla^{(h)}_a\mathcal{J}_{[3,2]}^a$ with $\mathcal{J}_{[3,2]}^a$ the most general linear combination of terms of the form
\begin{equation}
 A \nabla^{(h)}\nabla^{(h)}\nabla^{(h)}A\,,\quad
\nabla^{(h)}A\nabla^{(h)}\nabla^{(h)}A\,,\label{eq: J32}
\end{equation}
where we have left the indices unspecified with the understanding that all possible contractions are allowed. There are in total 15 possible none equivalent index contraction (leaving one index free) for each of the two terms in \eqref{eq: J32}. It turns out that the 15 terms made out of contractions of $A \nabla^{(h)}\nabla^{(h)}\nabla^{(h)}A$ all contribute to the on-shell action but are on-shell all linear combinations of the two terms obtained above in \eqref{eq: contribution BoxA^2} and \eqref{eq: contribution DivXh}. The 15 remaining terms made of contractions of the second term in \eqref{eq: J32} do not contribute at all on-shell.

All remaining terms in $\mathcal{T}_{[4,2]}$ modulo terms of the form $\sqrt{-h}\nabla^{(h)}_a\mathcal{J}_{[3,2]}^a$ are formed out all of possible contractions of terms of the form
\begin{equation}
\sqrt{-h}\nabla^{(h)}\nabla^{(h)}A\nabla^{(h)}\nabla^{(h)}A\,.\label{eq: T42}
\end{equation}
There are in total 12 such possible contractions. However, none of these terms contribute to the on-shell action.

Finally we note that extrinsic counterterms such as $F_{\mu\nu}n^\mu A^\nu$ and $S_{\mu\nu}n^\mu
A^\nu$ do contribute but not in a new and independent way from what we have found
already. We have not analyzed possible counterterms of order $\mathcal{T}_{[4,4]}$
and higher.

We conclude that we did not manage to holographically renormalize the action for the solutions of section \ref{subsec:propssolution} that have $U_{(4)}\neq 0$ using local counterterms as well as counterterms that are proportional to $\log r$. Had we compactified the $V$  coordinate and not considered a more general ansatz than \eqref{eq:ansatzAT}--\eqref{eq:ansatzgVV} we would have found that by demanding periodicity of the solution $U_{(4)}=0$ and the above problem with the holographic renormalization would be gone.
It would be very interesting to study more general ans\"atze with $V$ periodically identified and to see if we can holographically renormalize the action or not.

If we relax the condition that the non-locality of the counterterms is at most of $\log r$ type and nothing else then we can renormalize the action \eqref{eq: onshell action div} by adding local counterterms that are proportional to $\log^2 r$. In this light we note that on-shell the $r$ and $V$ dependencies are correlated as discussed in section \ref{subsec:commentsansatz}. The possibe need for non-local counterterms has also been discussed in \cite{Guica:2010sw}.

We would like to stress that without imposing any constraints we do not have a FG type expansion for the most general solution to the equations of motion with AlSch boundary conditions so any discussion of holographic renormalization will be constrained by the ansatz we make for the asymptotic expansions.

Apart from working with constraints there may be another possibility to obtain FG type expansions. FG type expansions could be obtained after Fourier transforming $V$ to say $m$ like we did in appendix \ref{app:scalars} (see also \cite{Guica:2010sw}). The main difference with appendix \ref{app:scalars} is that now we are trying to holographically reconstruct a
space-time metric. Choosing the radial $TV$ gauge and then Fourier transforming from $V$ to $m$ breaks space-time diffeomorphism covariance. Without fixing the free FG coefficients as functions of $m$ one can never undo the Fourier transformation, so also in this case in order to write something that is fully in configuration space one would have to impose constraints on the momentum $m$ dependence of the free FG coefficients which from a $V$ space point of view means that once again we must impose a constraint. Hence one either imposes a constraint and works in $(V,T,R)$ coordinates or one does not but then one is stuck in $(m,T,r)$-space forever.

At the beginning of this section we observed that for all known solutions for which a FG expansion does exist we can holographically renormalize the theory so perhaps apart from using AlSch boundary conditions we should also demand renormalizability by which we mean that for our Dirichlet boundary conditions there should exist a counterterm action that is local with the exception of a term that is proportional to $\log r$ (whose coefficient is again a local function on the cut-off boundary). It would be interesting to see if this would single out those solutions for which a FG expansion can be constructed.

\section{Discussion}\label{sec:discussion}

We have defined a notion of an AlSch space-time using a frame field decomposition and specific falloff conditions for the frame field. In a radial gauge the boundary value $A_{(0)}^a$ of the massive vector is proportional to the generator of particle number $N$. We do not impose any kind of symmetries on the asymptotic solution so $N$ is in general not an asymptotic Killing vector. We have shown that without imposing some kind of constraint on the full solution there does not exist a FG type expansion in the sense that without a constraint the solution has an intrinsic dependence on the ansatz made to write down the radial $r$ expansion of the metric and vector field and that even with a given ansatz there are new undetermined functions appearing at each order in the radial expansion. Both the problem of not being to able to identify the radial expansion and the problem of new functions appearing at higher orders are controlled by the role of particle number $N$ which in a suitable gauge, that we refer to as the radial $TV$ gauge is such that $N=\partial_V$ (with the Hamiltonian given by $\partial_T$). It is therefore the $V$ dependence that plays a central role in this work.

We have shown that this problem already occurs for the case of a free scalar field on a fixed Schr\"odinger background. There it is shown that this problem can be remedied in two ways: i). by imposing constraints on the $V$ dependence or ii). by Fourier transforming from $V$ to $m$ and working in $(m,T,r)$ space. It is only when we form superpositions of $m$-dependent solutions that it is crucial that we decide what the dependence of the FG coefficients is on $m$ in order to even in principle be able to perform the $m$ integral back to configuration $V$ space. This is the momentum space counterpart of the indeterminacy of the radial expansion for the most general solution. If we wish to write down a FG type expansion in $(V,T,r)$ space then there is only one option and that is to impose a constraint. If we do not mind working in $(m,T,r)$ space without ever going back to $(V,T,r)$ space then no constraints are needed. Even though for fields on a fixed Schr\"odinger background working in $(m,T,r)$ space seems a natural thing to do. It is less clear whether the same applies to studying AlSch space-times. In any case in this work we have chosen to look at AlSch space-times from a $(V,T,r)$ space perspective. It would be useful to further investigate the possibility of Fourier transforming $V$ in the context of constructing AlSch space-times. In this context it would be interesting to better understand the linearized perturbations of the metric and vector on a fixed Schr\"odinger background (see \cite{Guica:2010sw,vanRees:2012cw} for work in this direction).

Not much is known about the field theories dual to the Schr\"odinger backgrounds. They are expected to be dipole deformations of gauge theories \cite{Maldacena:2008wh} and in the context of 3D Schr\"odinger space-times they are referred to as dipole CFTs \cite{Song:2011sr}. In the case of a Schr\"odinger space-time we can expect on symmetry grounds that these theories organize themselves as a non-relativistic CFT. From the analysis of \cite{Nishida:2007pj} we know that primary operators (as defined in \cite{Nishida:2007pj}) are always in an eigenstate of particle number and hence have a fixed $m$-dependence. It would be very useful to understand better from the point of view of a non-relativistic CFT how particle number can be broken.

We have shown that imposing the constraint $g_{\mu\nu}A^\mu A^\nu=0$ leads to an exact solution (the FG series terminates) which is such that $\partial_V$ is a null Killing vector.
We have then relaxed the property that $\partial_V$ is null and Killing to it being either Killing or null. In the latter case we obtained FG type expansions for solutions that break particle number either in the UV or in the IR.

For each of these constrained cases one can write down a counterterm action that makes the on-shell action finite. However we have not been able to holographically renormalize the on-shell action for the solutions of section \ref{subsec:propssolution} that are only constrained by the ansatz \eqref{eq:ansatzAT}--\eqref{eq:ansatzgVV}. This problem disappeared by compactifying the $V$ coordinate. It could be insightful to study more general ans\"atze for the case of a compact $V$ coordinate in relation to the problem of holographic renormalization. So far we have the situation that for all solutions for which we are able to write down FG expansions we are able to holographically renormalize the theory and that whenever the FG expansion breaks down so too does the renormalizability. It would be interesting to explore this further.

Our frame field formulation of an AlSch space-time, which in radial gauge reads,
\begin{eqnarray}
e^+_a & = & r^{-2}\left(e^+_{(0)a}+o(1)\right)\,,\\
e^-_a & = & e^-_{(0)a}+o(1)\,,
\end{eqnarray}
is a natural starting point to define the boundary stress tensor by varying the on-shell action with respect to $e^+_{(0)a}$ and $e^-_{(0)a}$ following observations in \cite{Ross:2009ar,Ross:2011gu}. We know from the case of AlLif space-times with $z=2$ that some of the sources of the boundary stress-energy tensor are zero by definition of the $z=2$ AlLif
space-time so that one must include all irrelevant deformations that continuously deform the AlLif space-time in order to compute the boundary stress-energy tensor \cite{Ross:2011gu}. In the case of AlSch space-times the situation is more complex because
in $(V,T,r)$ space we only have FG expansions when we impose constraints and so we may need to be more careful when defining the boundary stress tensor. It would be interesting to study these and related problems further.

A closely related space-time to the Schr\"odinger space-time is spacelike warped AdS$_3$ of which null warped AdS$_3$ can be obtained via a scaling limit. It might therefore be insightful to try to define asymptotically spacelike warped AdS$_3$ space-times and to see if similar problems regarding the construction of FG expansions appear there as we found here.

\section*{Acknowledgements}

We express our gratitude to Matthias Blau for useful discussions and careful reading of this manuscript. The work of JH is supported by the Danish National Research Foundation
project ``Black holes and their role in quantum gravity''. BR acknowledges support from the Rosenfeld
fund of the Niels Bohr Institute, the Max Planck Research Fellowship and the Swiss National Science Foundation through the fellowship PBBEP2\_144805.
BR expresses his gratitude to the Niels Bohr Institute and the Max Planck Institute for Gravitational
Physics (AEI) in Potsdam for warm hospitality during various stages of this project.

Some of the calculations in this paper have been performed using the software
package Cadabra \cite{Peeters:2006kp,Peeters:2007wn}.

\appendix

\section{Locally Schr\"odinger space-times}\label{app:locallySch}

We review the properties of locally Schr\"odinger space-times. Some of the material here can also be found in \cite{Hartong:2012sw}.

\subsection{Defining a locally Schr\"odinger space-time}\label{subsec:Sch-space-time}

It is well-known that a $(d+3)$-dimensional $z=2$ Schr\"odinger space-time in string theory can be obtained by starting with an AdS$_{d+3}\times Y^{7-d}$ solution and performing a TsT transformation (or null Melvin twist) on it with the shift along a null Killing direction of the AdS$_{d+3}$ space-time and the T-duality along a circle in the compact internal space $Y^{7-d}$ \cite{Maldacena:2008wh,Herzog:2008wg,Adams:2008wt,Imeroni:2009cs,Hartong:2010ec,Bobev:2011qx,Detournay:2012dz}. To actually perform the TsT transformation requires adapted coordinates that make the two isometries (forming a 2-torus) involved manifest\footnote{The TsT transformation, as a solution generating technique, also applies when the null Killing direction is non-compact.}.

It is however possible to formulate the result of such a TsT transformation in a coordinate independent manner as follows. Let $\gamma_{\mu\nu}$ be the metric of an AdS$_{d+3}$ space-time and let $A^\mu$ be any null Killing vector field of $\gamma_{\mu\nu}$. Then the metric $g_{\mu\nu}$ defined as\footnote{One can also take the plus sign in the definition of $g_{\mu\nu}$, i.e. write $g_{\mu\nu}=\gamma_{\mu\nu}+A_\mu A_\nu$. However this metric violates the null energy condition so we will not consider it.}:
\begin{equation}\label{eq:Schmetric}
 g_{\mu\nu}=\gamma_{\mu\nu}-A_\mu A_\nu\,,
\end{equation}
is the metric of a $z=2$ Schr\"odinger space-time. The metric \eqref{eq:Schmetric} is of a generalized Kerr--Schild form. Note that $A_\mu=\gamma_{\mu\nu}A^{\nu}=g_{\mu\nu}A^{\nu}$.

We will now prove that $g_{\mu\nu}$ in \eqref{eq:Schmetric} locally describes a Schr\"odinger space-time. Let $K^\mu$ be a vector field defined on the AdS and Schr\"odinger space-times. We have
\begin{equation}
\mathcal{L}_K g_{\mu\nu}=\mathcal{L}_K\gamma_{\mu\nu}-A_\mu\mathcal{L}_KA_\nu-A_\nu\mathcal{L}_KA_\mu\,,
\end{equation}
where $\mathcal{L}_K$ denotes the Lie derivative along $K$. If
\begin{equation}
\mathcal{L}_K\gamma_{\mu\nu}=0\qquad\text{and}\qquad [K,A]=0\,,
\end{equation}
then $K$ is a Killing vector of $g_{\mu\nu}$. Since the commutator of any AdS null Killing vector is the Schr\"odinger algebra the isometry group of $g_{\mu\nu}$ contains the Schr\"odinger algebra. To prove that the isometry group of $g_{\mu\nu}$ does not contain more generators than the Schr\"odinger algebra we need to show that on AdS the only solutions to the equation
\begin{equation}
\mathcal{L}_K\gamma_{\mu\nu}-A_\mu\mathcal{L}_KA_\nu-A_\nu\mathcal{L}_KA_\mu=0\,,
\end{equation}
are those for which $\mathcal{L}_KA^\mu=0$. This follows from the fact that a $(d+3)$-dimensional space-time that admits at least the Schr\"odinger group as its isometries is either AdS or a Schr\"odinger space-time \cite{SchaferNameki:2009xr}.

We will shortly consider an alternative definition that allows us to define locally Schr\"odinger space-times by looking at the Riemann tensor. To this end we will make use of the fact that $A^\mu$ is hypersurface orthogonal with respect to both the AdS and the Schr\"odinger metric. To prove this we observe that on AdS $A^\mu$ satisfies the properties:
\begin{eqnarray}
0 & = & A^\mu A_\mu\,,\\
0 & = & \nabla^{(\gamma)}_\mu A_\nu+\nabla^{(\gamma)}_\nu A_\mu\,,\\
0 & = & A^\mu A^\nu R^{(\gamma)}_{\mu\nu}\,.
\end{eqnarray}
It follows that $A^\mu$ is tangent to a null geodesic congruence. Define
\begin{equation}
B_{\mu\nu}=\nabla^{(\gamma)}_\mu A_\nu\,.
\end{equation}
Further, introduce a second null vector $N^\mu$ satisfying
\begin{equation}
A^\mu N_\mu=-1\,.
\end{equation}
Define the projector
\begin{equation}
h_{\mu\nu}=\gamma_{\mu\nu}+A_\mu N_\nu+A_\nu N_\mu\,.
\end{equation}
This projects onto the co-dimension two subspace orthogonal to both $A^\mu$ and $N^\mu$. This space is not uniquely defined as $N^\mu$ is not uniquely defined. Anyway, the results will not depend on the specific choice for $N^\mu$. Next we define
\begin{equation}
\hat B_{\mu\nu}=h_\mu{}^\rho h_\nu{}^\sigma B_{\rho\sigma}\,.
\end{equation}
Because $A^\mu$ is Killing, the shear and expansion of the null geodesic congruence are zero and the Raychaudhuri equation reads
\begin{equation}
A^\mu A^\nu R^{(\gamma)}_{\mu\nu}=\hat\omega_{\mu\nu}\hat\omega^{\mu\nu}\,,
\end{equation}
where
\begin{equation}
\hat\omega_{\mu\nu}=\hat B_{[\mu\nu]}\,.
\end{equation}
Hence for Einstein space-times (such as AdS) we have for a null Killing vector that
\begin{equation}
\hat\omega_{\mu\nu}\hat\omega^{\mu\nu}=0\quad\rightarrow\quad\hat\omega_{\mu\nu}=0\,.
\end{equation}
We now show that this implies that $A^\mu$ is hypersurface orthogonal. We have from the definition of $\hat\omega_{\mu\nu}$ and the properties of $A^\mu$ that
\begin{equation}
0=\hat\omega_{\mu\nu}=\omega_{\mu\nu}+\omega_{\mu\rho}N^\rho A_\nu-\omega_{\nu\rho}N^\rho A_\mu\,,
\end{equation}
with $\omega_{\mu\nu}=\nabla^{(\gamma)}_{[\mu}A_{\nu]}$. In other words
\begin{equation}
\omega_{\mu\nu}=A_{[\mu}V_{\nu]}\,,
\end{equation}
where $V_\mu=2\omega_{\mu\rho}N^\rho$. It follows that for arbitrary $V_\mu$ and hence for any choice of $N^\mu$ that
\begin{equation}\label{eq:AisHSO}
A_{[\mu}\omega_{\nu\rho]}=A_{[\mu}\nabla^{(\gamma)}_\nu A_{\rho]}=0\,,
\end{equation}
which is the Frobenius integrability condition for $A^\mu$ to be hypersurface orthogonal. We conclude that any AdS null Killing vector is automatically hypersurface orthogonal. Using that the Christoffel symbols are related via
\begin{equation}\label{eq:barGamma}
 \bar\Gamma^\rho_{\mu\nu}=\Gamma^\rho_{\mu\nu}+\nabla^{(\gamma)\rho}(A_\mu A_\nu)\,,
\end{equation}
it follows that
\begin{equation}\label{eq:AisHSOonSch}
A_{[\mu}\nabla_\nu A_{\rho]}=A_{[\mu}\nabla^{(\gamma)}_\nu A_{\rho]}=0\,,
\end{equation}
so that $A^\mu$ is also hypersurface orthogonal with respect to the Schr\"odinger metric.

We are now ready to define a Schr\"odinger space-time via its Riemann tensor together with the existence of a null Killing vector $A^\mu$. To this end we start by computing the Riemann tensor of a pure Schr\"odinger space-time. Using \eqref{eq:barGamma} we compute the Riemann tensor of \eqref{eq:Schmetric} and using that the AdS Riemann tensor is given by \begin{equation}\label{eq:locallyAdS}
 R^{(\gamma)}_{\mu\nu\rho\sigma}=-\gamma_{\mu\rho}\gamma_{\nu\sigma}+\gamma_{\mu\sigma}\gamma_{\nu\rho}
\end{equation}
gives
\begin{eqnarray}
R_{\mu\nu\rho\sigma} & = & -g_{\mu\rho}g_{\nu\sigma}+g_{\mu\sigma}g_{\nu\rho}+g_{\mu\rho}A_\nu A_\sigma-g_{\nu\rho}A_\mu A_\sigma-g_{\mu\sigma}A_\nu A_\rho+
g_{\nu\sigma}A_\mu A_\rho\nonumber\\
&&+\frac{3}{4}F_{\mu\nu}F_{\rho\sigma}\,.\label{eq:SchRiemann}
\end{eqnarray}
To derive this form of the Riemann tensor we used that $\gamma_{\mu\nu}$ is an AdS metric and that $A^\mu$ is an AdS null Killing vector which is therefore hypersurface orthogonal, i.e. it satisfies \eqref{eq:AisHSO}. This form of the Riemann tensor supplied with the statement that $A^\mu$ is a null Killing vector of the metric $g_{\mu\nu}$ can be viewed as an alternative definition of a Schr\"odinger space-time. To prove this all one needs to do is to show that by starting from \eqref{eq:SchRiemann} the Riemann tensor of $\gamma_{\mu\nu}=g_{\mu\nu}+A_\mu A_\nu$ satisfies \eqref{eq:locallyAdS}, so that we end up with the decomposition \eqref{eq:Schmetric}. For this purpose it is useful to note that $A^\mu$ is also hypersurface orthogonal with respect to $g_{\mu\nu}$ as follows from \eqref{eq:AisHSOonSch}. This definition allows us to introduce the notion of a locally Schr\"odinger space-time, in analogy with locally AdS space-times, as those metrics satisfying \eqref{eq:SchRiemann} as well as admitting (locally) a null Killing vector. For later purposes we note that
\begin{equation}\label{eq:SchRiemanncontractedA}
 R_{\mu\nu\rho\sigma}A^\sigma=\left(-\gamma_{\mu\rho}\gamma_{\nu\sigma}+\gamma_{\mu\sigma}\gamma_{\nu\rho}\right)A^\sigma\,,
\end{equation}
i.e. upon contraction with $A^\mu$ the Schr\"odinger Riemann tensor behaves as an AdS Riemann tensor.

\subsection{The boundary}\label{subsec:boundary}

Consider a conformal rescaling of the AdS metric $\gamma_{\mu\nu}$ so that we get $\gamma'_{\mu\nu}=\Omega^2\gamma_{\mu\nu}$ where $\Omega$ is an AdS defining function. The vector field $A^\mu$ does not transform. If we apply these rescalings to the decomposition of the Schr\"odinger metric \eqref{eq:Schmetric} we obtain
\begin{equation}\label{eq:anisotropic rescaling}
g_{\mu\nu}=\Omega^{-2}\gamma'_{\mu\nu}-\Omega^{-4}A'_\mu A'_\nu\,,
\end{equation}
where $A'_\mu=\gamma'_{\mu\nu}A^\nu$. Let $\Omega=0$ denote the location of the AdS boundary. We write $g_{\mu\nu}=\Omega^{-2}g'_{\mu\nu}$ where
$g'_{\mu\nu}=\gamma'_{\mu\nu}-\Omega^{-2}A'_\mu A'_\nu$ then at leading order in a near boundary expansion we find
\begin{equation}
 R_{\mu\nu\rho\sigma}=-\Omega^{-4}g^{\kappa\tau}\frac{\partial_\kappa\Omega}{\Omega}\frac{\partial_\tau\Omega}{\Omega}\left(g'_{\mu\rho}g'_{\nu\sigma}-g'_{\mu\sigma}g'_{\nu\rho}\right)+O\left(\Omega^{-3}\right)\,,
\end{equation}
where the coefficients are functions of $g'_{\mu\nu}$. Contracting this with $A^\sigma$ and equating to \eqref{eq:SchRiemanncontractedA} we derive that
\begin{equation}
g^{\mu\nu}\frac{\partial_\mu\Omega}{\Omega}\frac{\partial_\nu\Omega}{\Omega}\vert_{\Omega=0}=1\,.
\end{equation}
Since $\gamma_{\mu\nu}$ is AdS and $\Omega$ an AdS defining function implying
\begin{equation}
\gamma^{\mu\nu}\frac{\partial_\mu\Omega}{\Omega}\frac{\partial_\nu\Omega}{\Omega}\vert_{\Omega=0}=1\,.
\end{equation}
It follows that
\begin{equation}\label{eq:Atangentialtobdry}
g^{\mu\nu}\frac{\partial_\mu\Omega}{\Omega}\frac{\partial_\nu\Omega}{\Omega}\vert_{\Omega=0}-\gamma^{\mu\nu}\frac{\partial_\mu\Omega}{\Omega}\frac{\partial_\nu\Omega}{\Omega}\vert_{\Omega=0}=\left(A^\mu\frac{\partial_\mu\Omega}{\Omega}\right)^2\vert_{\Omega=0}=0\,.
\end{equation}
We conclude that the Schr\"odinger defining function must satisfy the properties of an AdS defining function plus \eqref{eq:Atangentialtobdry} which states that $A^\mu$ must be tangential to the boundary as $\Omega^{-1}\partial_\mu\Omega$ is the unit normal.

We derived \eqref{eq:Atangentialtobdry} from boundary conditions for locally Schr\"odinger space-times. It can also be seen to arise by writing the locally Schr\"odinger metric in radial gauge, i.e. as
\begin{eqnarray}
 ds^2 & = & g_{\mu\nu}dx^\mu dx^\nu=\frac{dr^2}{r^2}+g_{ab}(r,x)dx^adx^b\,,\label{eq:Schradialgauge}\\
 A & = & A_r(r,x)dr+A_a(r,x)dx^a\,.
\end{eqnarray}
The metric \eqref{eq:Schradialgauge} will be a Schr\"odinger space-time if and only if $A_\mu$ is a null Killing vector of the AdS metric (or the Schr\"odinger metric for that matter) and the shifted metric
\begin{equation}\label{eq:AdSforradialSch}
 ds^2=\gamma_{\mu\nu}dx^\mu dx^\nu=(\frac{1}{r^2}+A_r^2)dr^2+2A_rA_adx^adr+\gamma_{ab}dx^adx^b\,,
\end{equation}
is AdS, where we defined $\gamma_{ab}=g_{ab}+A_aA_b$. Demanding that  \eqref{eq:AdSforradialSch} is AdS requires the Riemann tensor of $\gamma_{\mu\nu}$ to satisfy \eqref{eq:locallyAdS}. From the $rr$ component of the Killing equations for $A^\mu$ expressed in terms of $g_{\mu\nu}$ we derive
\begin{equation}
 A_r=\frac{A_{(0)r}(x)}{r}\,.
\end{equation}
The $ra$ term in \eqref{eq:AdSforradialSch} is of order $r^{-3}$ when $A_{(0)r}\neq 0$ since $A_{a}$ is of order $r^{-2}$. Hence it follows that $r=0$ is not an AdS boundary unless we put
\begin{equation}
A_{(0)r}=0\,,
\end{equation}
which is precisely the condition \eqref{eq:Atangentialtobdry}.

\subsection{FG expansions}\label{eq:FGexpansions}

We discuss here the construction of Fefferman--Graham expansions for 3-dimensional locally Schr\"odinger space-times. The higher-dimensional locally Schr\"odinger space-times were studied in \cite{Hartong:2012sw}. According to the definition of a locally Schr\"odinger space-time given in section \ref{subsec:Sch-space-time} we can employ the well-known Fefferman--Graham expansion for locally 3-dimensional AdS space-times together with the most general solution for a null Killing vector in such an AdS coordinate system.

Consider locally AdS$_{3}$ space-times. The complete solution to the equations defining a locally AdS$_{3}$ space-time \eqref{eq:locallyAdS} in FG gauge is given by \cite{Skenderis:1999nb}
\begin{equation}\label{eq:locallyAdSinFGcoord}
ds^2=\frac{dr^2}{r^2}+\gamma_{ab}dx^adx^b=\frac{dr^2}{r^2}+\frac{1}{r^2}\left(\gamma_{(0)ab}+r^2\gamma_{(2)ab}+r^4\gamma_{(4)ab}\right)dx^a dx^b\,,
\end{equation}
where $\gamma_{(2)ab}$ satisfies
\begin{eqnarray}
R_{(0)abcd} & = & -\gamma_{(0)ac}\gamma_{(2)bd}+\gamma_{(0)ad}\gamma_{(2)bc}+\gamma_{(0)bc}\gamma_{(2)ad}-\gamma_{(0)bd}\gamma_{(2)ac}\,,\label{eq:confflatness1}\\
0 & = & \nabla^{(0)}_{a}\gamma_{(2)bc}-\nabla^{(0)}_b\gamma_{(2)ac}\,,\label{eq:confflatness2}
\end{eqnarray}
and where $\gamma_{(4)ab}$ is given by
\begin{equation}
 \gamma_{(4)ab} = \frac{1}{4}\gamma_{(2)a}{}^c\gamma_{(2)cb}\,.
\end{equation}
Indices are raised and lowered with respect to the metric $\gamma_{(0)ab}$. For 2-dimensional boundaries the solution can be written as \cite{Skenderis:1999nb}
\begin{eqnarray}
 \gamma_{(2)ab} & = & -\frac{1}{2}R_{(0)}\gamma_{(0)ab}+t_{ab}\,,\\
 \nabla^{(0)a}t_{ab} & = & 0\,,\\
 t^a{}_a & = & \frac{1}{2}R_{(0)}\,,\label{eq:traceT}
\end{eqnarray}
where $t_{ab}$ is the boundary stress tensor. The inverse metric can be found from
\begin{equation}\label{eq:inverseh}
\left(\delta_a^c+\frac{r^2}{2}\gamma_{(2)}{}^c{}_a\right)\gamma_{(0)cd}\left(\delta^d_b+\frac{r^2}{2}\gamma_{(2)}{}^d{}_b\right)=r^2\gamma_{ab}\,,
\end{equation}
and the Christoffel symbols of the metric $\gamma_{ab}$ can be found from
\begin{equation}\label{eq:Christoffel}
\left(\delta_d^c+\frac{r^2}{2}\gamma_{(2)}{}^c{}_d\right)\Gamma^{(\gamma)}{}^d_{ab}=\Gamma^{(0)}{}^c_{ab}+\frac{r^2}{2}\nabla^{(0)}_b\gamma_{(2)}{}^c{}_a\,.
\end{equation}

Let $A^\mu$ denote any AdS null Killing vector. The AdS Killing equations
\begin{equation}\label{eq:KE}
 \nabla^{(\gamma)}_\mu A_\nu+\nabla^{(\gamma)}_\nu A_\mu=0
\end{equation}
are solved by
\begin{eqnarray}
A_r & = & \frac{A_{(0)r}}{r}\,,\label{eq:solutionKE1}\\
A_a & = & A^{b}_{(0)}\gamma_{ab}-\frac{1}{2}\partial_aA_{(0)r}-\frac{r^2}{4}\gamma_{(2)a}{}^b\partial_bA_{(0)r}\,,\label{eq:solutionKE2}
\end{eqnarray}
where $A_{(0)}^a$ is a conformal Killing vector of $\gamma_{(0)ab}$, i.e. a solution to
\begin{equation}
\nabla^{(0)}_aA_{(0)b}+\nabla^{(0)}_bA_{(0)a}=2A_{(0)r}\gamma_{(0)ab}\,,\label{eq:solutionKE3}
\end{equation}
from which we read off that $A_{(0)r}$ is given by
\begin{equation}\label{eq:defK}
A_{(0)r}=\frac{1}{2}\nabla^{(0)}_aA^a_{(0)}\,.
\end{equation}
Equation \eqref{eq:solutionKE1} follows from the $rr$ component of \eqref{eq:KE} and equation \eqref{eq:solutionKE2} follows from the $ar$ component of \eqref{eq:KE} in which we used \eqref{eq:inverseh}. The $ab$ component of the Killing equations at lowest order gives \eqref{eq:solutionKE3} and at next-to-leading order gives
\begin{equation}
 \mathcal{L}_{A_{(0)}}\gamma_{(2)ab}=\nabla^{(0)}_a\nabla^{(0)}_bA_{(0)r}\,,
\end{equation}
where $\mathcal{L}_{A_{(0)}}$ denotes the Lie derivative along the conformal Killing vector $A_{(0)}^a$. It can rewritten as the following transformation rule for $t_{ab}$
\begin{equation}\label{eq:improvement}
 \mathcal{L}_{A_{(0)}}t_{ab}=\left(\nabla^{(0)}_a\nabla^{(0)}_b-\gamma_{(0)ab}\square^{(0)}\right)A_{(0)r}\,,
\end{equation}
expressing that the Lie derivative of the boundary stress tensor along a conformal Killing direction is an improvement transformation. The next-to-next-to-leading order term is automatically satisfied and there are no higher order terms.

Imposing that $A^\mu$ is null everywhere in the bulk implies that we need to satisfy on top of $A_{(0)}^a$ being null the equations
\begin{eqnarray}
 A_{(0)}^aA_{(0)}^bT_{ab} & = & A_{(0)}^a\partial_a A_{(0)r}-A_{(0)r}^2\,,\\
 A_{(0)}^aT_{ac}A_{(0)}^bT_{b}{}^c & = & 2A_{(0)}^aT_{a}{}^b\partial_bA_{(0)r}-\partial_aA_{(0)r}\partial^aA_{(0)r}-R_{(0)}A_{(0)r}^2\,.
\end{eqnarray}

If we put $A_{(0)r}=0$ the last two equations together with the fact that $A_{(0)}^aA_{(0)}^b\gamma_{(0)ab}=0$ tell us that $A_{(0)}^aT_{ab}$ is proportional to $A_{(0)b}$. Using furthermore the fact that $\nabla^{(0)}_a A_{(0)}^a=0$ and the vanishing of the Einstein tensor we have $0=4R_{(0)ab}A_{(0)}^aA_{(0)}^b=F_{(0)}^2-S_{(0)}^2=F_{(0)}^2$ implying that $F_{(0)ab}=0$. From this it furthermore follows that $\square^{(0)}A_{(0)}^a=0$. Then using the Killing identity we find that $R_{(0)}=0$ so that according to \eqref{eq:traceT} $T_{ab}$ must be traceless. This can only be the case if we have
\begin{equation}
T_{ab}=a_{(2)}A_{(0)a}A_{(0)b}
\end{equation}
where $\mathcal{L}_{A_{(0)}}a_{(2)}=0$ as follows from \eqref{eq:improvement}.

Hence in an AdS radial gauge with $A_{(0)r}=rA_r\vert_{r=0}=0$ we obtain
\begin{eqnarray}
ds^2 & = & \frac{dr^2}{r^2}+g_{ab}dx^a dx^b\,,\\
g_{ab} & = & -\frac{1}{r^4}A_{(0)a}A_{(0)b}+\frac{1}{r^2}\gamma_{(0)ab}+a_{(2)}A_{(0)a}A_{(0)b}\,,\\
A^a & = & A^a_{(0)}\,,\\
A^r & = & 0\,,
\end{eqnarray}
in which $\gamma_{(0)ab}$ is flat and $A_{(0)}^a$ is a boundary null KV
\begin{eqnarray}
\nabla^{(0)}_aA_{(0)b}+\nabla^{(0)}_bA_{(0)a} & = & 0\,,\\
A_{(0)}^aA_{(0)}^b\gamma_{(0)ab} & = & 0\,.
\end{eqnarray}

\section{Expansions of the equations of motion}\label{app:solutions}

In this appendix we collect some background information on the derivation of the solutions presented in sections \ref{subsec:A^2=0solution} and \ref{subsec:propssolution}.

\subsection{$A^\mu$ is null: expansions for $\gamma_{\mu\nu}$ and $A^\mu$}\label{subsection:thesolutions}

In order to solve the equations of motion \eqref{eq:Einsteineqs-AdSside-A2=0-v2} and \eqref{eq:vectoreq1-AdSside-A2=0-v2} and constraints \eqref{eq:constraintA2=0}, \eqref{eq:vectoreq2-AdSside-A2=0} we will proceed by making an ansatz for the subleading terms in \eqref{eq:ansatz1}-\eqref{eq:ansatz4}.

We will consider a solution space made of all solutions which are of polynomial form with increasing powers of logs. More precisely we consider an ansatz of the form
\begin{eqnarray}
\gamma_{ab} & = & \frac{1}{r^2}\sum^\infty_{n=0}\sum^n_{m=0}r^{2n}\log^m r \gamma_{(2n,m)ab}\,,\label{eq:habto1storder}\\
A^a & = & \sum^\infty_{n=0}\sum^n_{m=0}r^{2n}\log^m r A^a_{(2n,m)}\,,\label{eq:Aato1storder}\\
A^r & = & r\sum^\infty_{n=1}\sum^n_{m=0}r^{2n}\log^m r A^r_{(2n,m)}\,,\label{eq:Arto1storder}
\end{eqnarray}
where each $\gamma_{(2n,m)ab}$, $A^a_{(2n,m)}$ and $A^r_{(2n,m)}$ are fully arbitrary\footnote{We denote $\gamma_{(2n,0)ab}$ by $\gamma_{(2n)ab}$ and similarly for the other coefficients.}. Indices are all raised and lowered with respect to $\gamma_{(0)ab}$. The properties of the leading order have already been given in section \ref{subsubsec:boundaryconditions}, here we solve for the subleading terms. For the purpose of clarity we will present explicitly how the equations of motion and constraints are solved without including the logs. However, it is straightforward to include them and the final result is the same in both cases as for the case with the $g_{\mu\nu}A^\mu A^\nu=0$ constraint these terms can be shown not to contribute.

We will employ a null-bein basis of our 2-dimensional boundary metric by writing
\begin{equation}\label{bnd: decompos metric}
\gamma_{(0)ab}=-A_{(0)a}N_{(0)b}-A_{(0)b}N_{(0)a}\,,
\end{equation}
where $N^a_{(0)}$ is an arbitrary vector field satisfying $N_{(0)a}N^a_{(0)}=0$ and $N_{(0)a}A^a_{(0)}=-1$ and $A_{(0)}^a$ is the boundary value of $A^a$. It follows that all quantities can be decomposed according to
\begin{eqnarray}
\gamma_{(2n)ab} &=& a_{(2n)}A_{(0)a}A_{(0)b} + b_{(2n)}N_{(0)a}N_{(0)b} + c_{(2n)}\gamma_{(0)ab}\,,\\
A_{(2n)}^a &=& \alpha_{(2n)}A^a_{(0)} + \beta_{(2n)}N^a_{(0)}\,,
\end{eqnarray}
where $a_{(2n)}, b_{(2n)}, c_{(2n)}, \alpha_{(2n)}, \beta_{(2n)}$ are fully arbitrary functions of the boundary coordinates. Using that we have $A_{(0)a}A^a_{(0)}=0$ and $\nabla_{(0)a}A^a_{(0)}=0$ it follows that the covariant derivative acting on $A_{(0)a}$ can be expressed as
\begin{equation}\label{eq:defsigma}
\nabla_{(0)a}A_{(0)b} = \sigma_{(0)} A_{(0)a}A_{(0)b}\,,
\end{equation}
for some arbitrary function $\sigma_{(0)}$.

We now proceed with the actual expansion of the equations of motion \eqref{eq:Einsteineqs-AdSside-A2=0-v2}, \eqref{eq:vectoreq1-AdSside-A2=0-v2} and constraints \eqref{eq:constraintA2=0}, \eqref{eq:vectoreq2-AdSside-A2=0} and work out their consequences order by order. The equations of motion and constraints vanish automatically at the following leading orders
\begin{eqnarray}
 \mbox{Eins}_{ab} \; @\; r^{-4}  & = & 0 \,,\\
 \mbox{Eins}_{ar} \; @\; r^{-3}  & = & 0 \,,\\
 \mbox{Eins}_{rr} \; @\; r^{-2}  & = & 0 \,,\\
 \mbox{Vec}_{a} \; @\; r^{-2} & = & 0 \,,\\
 \mbox{Vec}_{r} \; @\; r^{-1} & = & 0 \,,\\
 \nabla^{(\gamma)}_\mu A^\mu \; @\; r^0 & = & 0 \,,\\
 \gamma_{\mu\nu}A^\mu A^\nu \; @\; r^{-2} & = & 0 \,.
\end{eqnarray}
At the next order, denoting by $\mathcal{L}_{A_{(0)}}$ and $\mathcal{L}_{N_{(0)}}$ the Lie derivatives along $A^a_{(0)}$ and $N^a_{(0)}$ respectively, we find
\begin{eqnarray}
\mbox{Eins}_{ab} \; @\; r^{-2} & = & - \mathcal{L}_{A_{(0)}}\sigma_{(0)} A_{(0)a} A_{(0)b} - 4\alpha_{(2)} \beta_{(2)}  A_{(0)a} A_{(0)b}  \nonumber\\
&& + \sigma_{(0)}\mathcal{L}_{A_{(0)}}b_{(2)}  A_{(0)a} A_{(0)b}- 2\sigma_{(0)}  \mathcal{L}_{A_{(0)}}\beta_{(2)} A_{(0)a} A_{(0)b} \,,\\
\mbox{Eins}_{ar} \; @\; r^{-1} & = &- \mathcal{L}_{A_{(0)}}\alpha_{(2)} A_{(0)a} - \mathcal{L}_{A_{(0)}}\beta_{(2)} N_{(0)a} + \sigma_{(0)} \beta_{(2)}A_{(0)a}  \,,\\
\mbox{Eins}_{rr} \; @\; r^{0} & = & - 2 \mathcal{L}_{A_{(0)}}A^r_{(2)} -\sigma_{(0)}  \mathcal{L}_{A_{(0)}}b_{(2)} + 2\sigma_{(0)} \mathcal{L}_{A_{(0)}}\beta_{(2)}  + 8 \alpha_{(2)} \beta_{(2)} \,,\\
\mbox{Vec}_{a} \; @\; r^{0} & = & - 4 A_{(0)a} \alpha_{(2)} + 2\mathcal{L}_{A_{(0)}}\sigma_{(0)} A_{(0)a} - 4 N_{(0)a} \beta_{(2)} + 8A_{(0)a} \alpha_{(2)} \beta_{(2)} \nonumber\\
&& - 2 \mathcal{L}_{A_{(0)}}b_{(2)} \sigma_{(0)} A_{(0)a} + 4 \mathcal{L}_{A_{(0)}}\beta_{(2)} \sigma_{(0)} A_{(0)a} \,,\\
\mbox{Vec}_{r} \; @\; r & = & - 4A^r_{(2)} + 2 \mathcal{L}_{A_{(0)}}c_{(2)} + 4\mathcal{L}_{A_{(0)}}\alpha_{(2)} + 4 \mathcal{L}_{N_{(0)}}\beta_{(2)} - 2\sigma_{(0)} b_{(2)} \nonumber\\
&&+ 4 \sigma_{(0)} \beta_{(2)} \,,\\
\nabla^{(\gamma)}_\mu A^\mu \; @\; r^2 & = & \mathcal{L}_{A_{(0)}}c_{(2)} + \mathcal{L}_{A_{(0)}}\alpha_{(2)}+ \mathcal{L}_{N_{(0)}}\beta_{(2)} + \sigma_{(0)} \beta_{(2)} \,,\\
\gamma_{\mu\nu}A^\mu A^\nu \; @\; r^{0} & = & b_{(2)} - 2\beta_{(2)} \,,
\end{eqnarray}
whose unique solution is
\begin{eqnarray}
0 & = & b_{(2)} =  \alpha_{(2)} =\beta_{(2)} = A^r_{(2)} \,,\\
0 & = & \mathcal{L}_{A_{(0)}}\sigma_{(0)} = \mathcal{L}_{A_{(0)}}c_{2}\,.
\end{eqnarray}
Here the condition $\mathcal{L}_{A_{(0)}}\sigma_{(0)}=0$ can be seen to be the equivalent of \eqref{eq:Y0ab=0}. By repeating the argument just below \eqref{eq:F0ab} using \eqref{eq:F0ab} and the fact that $\mathcal{L}_{A_{(0)}}\sigma_{(0)}=0$ we just found again that the boundary metric $\gamma_{(0)ab}$ is flat. Because $\alpha_{(2)}=\beta_{(2)}=0$ we also find that
\begin{equation}
A^a_{(2)}=0\,.
\end{equation}
Using this information we now solve the following set of equations
\begin{eqnarray}
\mbox{Eins}_{ab} \; @\; r^{0} & = & 2c_{(2)} h_{(0)ab} + \left( \sigma_{(0)}\mathcal{L}_{A_{(0)}}(b_{(4)} - 2 \beta_{(4)})\right.\nonumber\\
&&\left. - \frac{1}{2}\mathcal{L}_{A_{(0)}}\mathcal{L}_{A_{(0)}}a_{(2)} \right)A_{(0)a} A_{(0)b}\,,\\
A^a_{(0)}A^b_{(0)}\mbox{Eins}_{ab} \; @\; r^{2} & = & - \frac{1}{2} \mathcal{L}_{A_{(0)}}\mathcal{L}_{A_{(0)}}b_{(4)} + \mathcal{L}_{A_{(0)}}\mathcal{L}_{A_{(0)}}\beta_{(4)} - 4b_{(4)}\,,\\
\gamma_{\mu\nu} A^\mu A^\nu \; @\; r^{2} & = & b_{(4)} - 2 \beta_{(4)}\,,
\end{eqnarray}
from which we learn that
\begin{eqnarray}
0 & = & b_{(4)} = \beta_{(4)} = c_{(2)} \,,\\
0 & = & \mathcal{L}_{A_{(0)}}\mathcal{L}_{A_{(0)}}a_{(2)} \,.\label{eq: LALAa20}
\end{eqnarray}
The remaining equations become
\begin{eqnarray}
\mbox{Eins}_{ar} \; @\; r & = & A_{(0)a} \mathcal{L}_{A_{(0)}}a_{(2)} - 2 A_{(0)a} \mathcal{L}_{A_{(0)}}\alpha_{(4)} \,,\\
\mbox{Eins}_{rr} \; @\; r^{2} & =  & - 4 \mathcal{L}_{A_{(0)}}A^r_{(4)} - 8 c_{(4)}\,,\\
\mbox{Vec}_{a} \; @\; r^{2} & = & 0 \,,\\
\mbox{Vec}_{r} \; @\; r^3 & =  & 12 A^r_{(4)} + 2 \mathcal{L}_{A_{(0)}}c_{(4)} + 6\mathcal{L}_{A_{(0)}}\alpha_{(4)}\,,\\
\nabla^{(\gamma)}_\mu A^\mu \; @\; r^4 & = & 2 A^r_{(4)} + \mathcal{L}_{A_{(0)}}c_{(4)} + \mathcal{L}_{A_{(0)}}\alpha_{(4)} \,,
\end{eqnarray}
and using \eqref{eq: LALAa20} these equations are solved by
\begin{eqnarray}
c_{(4)} & = & 0\,,\\
\mathcal{L}_{A_{(0)}}a_{(2)} & = & 2\mathcal{L}_{A_{(0)}}\alpha_{(4)} \,,\label{eq:Liea2}\\
A^r_{(4)} & = & -\frac{1}{2}\mathcal{L}_{A_{(0)}}\alpha_{(4)}\,.\label{eq:Ar4}
\end{eqnarray}
Proceeding in the same way at the next order we get from
\begin{eqnarray}
A^a_{(0)}A^b_{(0)}\mbox{Eins}_{ab} \; @\; r^4 & = & - \frac{1}{2} \mathcal{L}_{A_{(0)}}\mathcal{L}_{A_{(0)}}b_{(6)} + \mathcal{L}_{A_{(0)}}\mathcal{L}_{A_{(0)}}\beta_{(6)} - 12 b_{(6)}\,,\\
\gamma_{\mu\nu}A^\mu A^\nu \; @\; r^4 & = & b_{(6)} - 2 \beta_{(6)} \,,
\end{eqnarray}
that
\begin{equation}
b_{(6)} = \beta_{(6)} = 0\,.
\end{equation}
Further, using this result we then find
\begin{eqnarray}
\mbox{Eins}_{ab} \; @\; r^2 & = & \left(- \frac{1}{2} \square^{(0)}\alpha_{(4)} - \frac{1}{2} \mathcal{L}_{A_{(0)}}\mathcal{L}_{A_{(0)}}a_{(4)} - 4 a_{(4)}\right) A_{(0)a}
A_{(0)b} \,,\\
\mbox{Eins}_{ar} \; @\; r^3 & = & - \frac{1}{4} A_{(0)a} \square^{(0)}A^r_{(4)} + 2A_{(0)a} \mathcal{L}_{A_{(0)}}a_{(4)} - 3A_{(0)a} \mathcal{L}_{A_{(0)}}\alpha_{(6)} \,,\\
\mbox{Eins}_{rr} \; @\; r^4 & = & - 6 \mathcal{L}_{A_{(0)}}A^r_{(6)} - 24c_{(6)} \,,\\
\mbox{Vec}_{a} \; @\; r^4 & = & \left(\frac{3}{2} \square^{(0)}\alpha_{(4)} +\mathcal{L}_{A_{(0)}}\mathcal{L}_{A_{(0)}}a_{(4)} + 12 \alpha_{(6)} - 6A^r_{(4)} \sigma_{(0)}\right)A_{(0)a}   \,,\\
\mbox{Vec}_{r} \; @\; r^5 & = & 44A^r_{(6)} + \square^{(0)}A^r_{(4)} + 2\mathcal{L}_{A_{(0)}}c_{(6)} + 8 \mathcal{L}_{A_{(0)}}\alpha_{(6)}\,,\\
\nabla^{(\gamma)}_\mu A^\mu \; @\; r^6 & = & 4 {A^r_{(6)}} + \mathcal{L}_{A_{(0)}}c_{(6)} + \mathcal{L}_{A_{(0)}}\alpha_{(6)} \,,
\end{eqnarray}
whose unique solution is
\begin{eqnarray}
\mathcal{L}_{A_{(0)}}\mathcal{L}_{A_{(0)}}a_{(4)}  & = &- 8a_{(4)} - \square^{(0)}\alpha_{(4)}\,,\\
c_{(6)} & = &- \frac{2}{9}a_{(4)} - \frac{1}{36}\square^{(0)}\alpha_{(4)} \,,\\
\alpha_{(6)}   & = & \frac{2}{3}a_{(4)}  +\frac{1}{2}\sigma_{(0)} A^r_{(4)} - \frac{1}{24}\square^{(0)}\alpha_{(4)} \,,\\
A^r_{(6)} & = &- \frac{1}{9}\mathcal{L}_{A_{(0)}}a_{(4)}\,,
\end{eqnarray}
where we used that $\mathcal{L}_{A_{(0)}}\square^{(0)}\phi_{(0)}=\square^{(0)}\mathcal{L}_{A_{(0)}}\phi_{(0)}=0$ for a function $\phi_{(0)}$ satisfying\linebreak $\mathcal{L}_{A_{(0)}}\mathcal{L}_{A_{(0)}}\phi_{(0)}=0$.
To prove this use \eqref{eq:defsigma} as well as $\nabla^{(0)}_aN_{(0)b}=-\sigma_{(0)}A_{(0)a}N_{(0)b}$. Next we deduce,
\begin{equation}
b_{(8)}=\beta_{(8)}=0\,,
\end{equation}
as before from
\begin{eqnarray}
A^a_{(0)}A^b_{(0)}\mbox{Eins}_{ab} \; @\; r^6 & = & - \frac{1}{2}\mathcal{L}_{A_{(0)}}\mathcal{L}_{A_{(0)}}b_{(8)} + \mathcal{L}_{A_{(0)}}\mathcal{L}_{A_{(0)}}\beta_{(8)} - 24 b_{(8)}\,,\\
\gamma_{\mu\nu}A^\mu A^\nu \; @\; r^6 & = & b_{(8)} - 2\, \beta_{(8)} \,.
\end{eqnarray}
To proceed we now first look only at
\begin{equation}
A^a_{(0)}\mbox{Vec}_{a} \; @\; r^6 = \mathcal{L}_{A_{(0)}}\mathcal{L}_{A_{(0)}}c_{(6)}  + \mathcal{L}_{A_{(0)}}\mathcal{L}_{A_{(0)}}\alpha_{(6)} - 8 c_{(6)} \,,
\end{equation}
from which we learn
\begin{equation}
\mathcal{L}_{A_{(0)}}\mathcal{L}_{A_{(0)}}a_{(4)} = 0 \quad \Rightarrow \quad c_{(6)} = A^r_{(6)} = 0\,.
\end{equation}
Using this information together with $\square^{(0)}\alpha_{(6)}=0$ which follows from $\mathcal{L}_{A_{(0)}}\alpha_{(6)}=0$ as a consequence of $\mathcal{L}_{A_{(0)}}a_{(4)}=0$ we get for the remaining equations
\begin{eqnarray}
\mbox{Eins}_{ab} \; @\; r^4 & = & \left( - \frac{1}{2}\mathcal{L}_{A_{(0)}}\mathcal{L}_{A_{(0)}}a_{(6)} + 36 (A^r_{(4)})^{2} + 8\alpha_{(4)}^{2} - 12 a_{(6)} \right)A_{(0)a}A_{(0)b} \,,\\
\mbox{Eins}_{ar} \; @\; r^5 & = & \left(3\mathcal{L}_{A_{(0)}}a_{(6)} - 4 \mathcal{L}_{A_{(0)}}\alpha_{(8)} + 8A^r_{(4)} \alpha_{(4)} \right) A_{(0)a}\,,\\
\mbox{Eins}_{rr} \; @\; r^6 & = & -8\mathcal{L}_{A_{(0)}}A^r_{(8)} - 40 (A^r_{(4)})^{2} - 48 c_{(8)} \,,\\
\mbox{Vec}_{a} \; @\; r^6 & = & \left(\mathcal{L}_{A_{(0)}}\mathcal{L}_{A_{(0)}}a_{(6)}- 56 (A^r_{(4)})^{2} + 32 \alpha_{(8)}\right)A_{(0)a}  \,,\\
\mbox{Vec}_{r} \; @\; r^7 & = & 92 A^r_{(8)} + 2 \mathcal{L}_{A_{(0)}}c_{(8)} + 10 \mathcal{L}_{A_{(0)}}\alpha_{(8)}\,,\\
\nabla^{(\gamma)}_\mu A^\mu \; @\; r^8 & = &6 A^r_{(8)} + \mathcal{L}_{A_{(0)}}c_{(8)} + \mathcal{L}_{A_{(0)}}\alpha_{(8)} \,,
\end{eqnarray}
which are solved by
\begin{eqnarray}
\mathcal{L}_{A_{(0)}}\mathcal{L}_{A_{(0)}}a_{(6)} & = & - 24a_{(6)} + 72(A^r_{(4)})^{2} + 16\alpha_{(4)}^2 \,,\\
c_{(8)} & =  & - \frac{3}{10}a_{(6)} + \frac{1}{5}\alpha_{(4)}^2 \,,\\
\alpha_{(8)} & = & \frac{3}{4}a_{(6)} - \frac{1}{2}(A^r_{(4)})^{2} - \frac{1}{2}\alpha_{(4)}^2\,,\\
A^r_{(8)} & = &- \frac{1}{10} \mathcal{L}_{A_{(0)}}\alpha_{(8)} \,.
\end{eqnarray}
Expanding now at order ten things slightly change and we obtain
\begin{eqnarray}
\gamma_{\mu\nu}A^\mu A^\nu \; @\; r^8 & = & (A^r_{(4)})^{2} + b_{(10)} - 2 \beta_{(10)} \,,\\
A^a_{(0)}A^b_{(0)}\mbox{Eins}_{ab} \; @\; r^8 & = &- \frac{1}{2} \mathcal{L}_{A_{(0)}}\mathcal{L}_{A_{(0)}}b_{(10)} + \mathcal{L}_{A_{(0)}}\mathcal{L}_{A_{(0)}}\beta_{(10)} - 40 b_{(10)}\,.
\end{eqnarray}
This time it follows that
\begin{equation}
b_{(10)} = 0 \,, \qquad \beta_{(10)}= \frac{1}{2}(A^r_{(4)})^{2} \,.
\end{equation}
We then use
\begin{equation}
A^a_{(0)}\mbox{Vec}_{a} \; @\; r^8 =  4 \mathcal{L}_{A_{(0)}}A^r_{(8)} + \mathcal{L}_{A_{(0)}}\mathcal{L}_{A_{(0)}}c_{(8)} + \mathcal{L}_{A_{(0)}}\mathcal{L}_{A_{(0)}}\alpha_{(8)} - 60\beta_{(10)}  \,,
\end{equation}
which can be seen to be equivalent to
\begin{equation}
a_{(6)} = \frac{2}{3}\alpha_{(4)}^2 - \frac{50}{9}(A^r_{(4)})^{2}\,.
\end{equation}
From that expression it is easy to conclude that (using \eqref{eq: LALAa20}, \eqref{eq:Liea2} and \eqref{eq:Ar4})
\begin{equation}
\mathcal{L}_{A_{(0)}}\mathcal{L}_{A_{(0)}}a_{(6)} = \frac{4}{3}\mathcal{L}_{A_{(0)}}\alpha_{(4)}^2\,,
\end{equation}
and hence it necessarily must be that
\begin{equation}
A^r_{(4)} = 0\,.
\end{equation}

It is quite remarkable that this information is hidden at the N$^5$LO (where by LO we mean the $n=0$ term in \eqref{eq:habto1storder}--\eqref{eq:Arto1storder}) in the expansion and implies an important collapse of the asymptotic expansion to an exact solution. Indeed, solving the equations of motion and constraints further we would no longer find any non-zero coefficients. However for some coefficients we would notice it with a two order delay. For example the coefficient $a_{(8)}$ is a N$^4$LO term but we would discover that it vanishes only by expanding up to N$^6$LO. As mentioned before working with the more general ansatz \eqref{eq:habto1storder}-\eqref{eq:Arto1storder}, i.e. including the logarithmic terms, would not have made any difference.

\subsection{The unconstrained case: properties of $g_{(0)ab}$}\label{subsec:propsofg0ab}

In this subsection we show that $e^{+}_{(2,1)a}$ and $e^{+}_{(2)a}$ are
proportional to $A_{(0)a}$ on-shell. In order to use the equations of motion we take
advantage of the fact that $\gamma_{(0)ab}$ is non-degenerate to introduce without
loss of generality the coordinates ($T,V$) together with the gauge
choice\footnote{We warn the reader that these $TV$ coordinates are not equivalent to
what we call the $TV$ gauge in subsection \ref{subsec:radialTVgauge}.}
\begin{equation}
\gamma_{(0)ab}dx^adx^b = 2H_{(0)}dTdV\,,
\end{equation}
for some arbitrary (non-vanishing) function $H_{(0)}=H_{(0)}(T,V)$. It then follows
from $A^a_{(0)}$ being null with respect to $\gamma_{(0)ab}$ that
$A^T_{(0)}A^V_{(0)}=0$. We choose $A^T_{(0)}=0$ implying $A_{(0)V}=0$ and
$A_{(0)T}=H_{(0)}A^V_{(0)}$. We write $A^V_{(0)}\equiv V_{(0)}$ and by assumption we
have $H_{(0)}\neq 0$ and $V_{(0)}\neq0$. It follows from the expansion
\eqref{eq:expansiong} and the equations \eqref{eq:g-2}-\eqref{eq: relation between
g0 and gam0} that
\begin{eqnarray}
g_{TT} &=& -\frac{1}{r^4}(H_{(0)}V_{(0)})^2 - \frac{2H_{(0)}V_{(0)}}{r^2}\left(\log
re^{+}_{(2,1)T} + e^{+}_{(2)T}\right) + O(\log^2 r)\,,\\
g_{TV} &=& \frac{H_{(0)}}{r^2}-\frac{H_{(0)}V_{(0)}}{r^2}\left(\log re^{+}_{(2,1)V}
+ e^{+}_{(2)V}\right) + \log^2 r g_{(2,2)TV}  \,,\nonumber\\
&& + \log r g_{(2,1)TV}+g_{(2)TV} + O(r^2\log^3 r)\,,\\
g_{VV} &=& \log^2r g_{(2,2)VV} + \log r g_{(2,1)VV}+g_{(2)VV} + O(r^2\log^3 r)\,.
\end{eqnarray}
The vector field read
\begin{eqnarray}
A^T &=& r^2(\log rA^T_{(2,1)}+A^T_{(2)})+ O(r^4\log^2r)\,,\\
A^V &=& V_{(0)} + r^2(\log r A^V_{(2,1)}+A^V_{(2)})+ O(r^4\log^2 r)\,,\\
A^r &=& r^3(\log r A^r_{(2,1)}+A^r_{(2)})+ O(r^5\log^2r)\,.
\end{eqnarray}
The boundary condition \eqref{eq:A constraint on AdS} imply the relations
\begin{eqnarray}
e^{+}_{(2,1)V} &=& -H_{(0)}A^T_{(2,1)}\,,\\
e^{+}_{(2)V} &=& -H_{(0)}A^T_{(2)}\,,
\end{eqnarray}
whereas \eqref{eq:Asquare constraint on AdS} gives
\begin{eqnarray}
g_{(2,2)VV} &=& -(H_{(0)}A^T_{(2,1)})^2\,,\\
g_{(2,1)VV} &=&
-\frac{2H_{(0)}A^T_{(2,1)}}{V_{(0)}}\left(H_{(0)}V_{(0)}A^T_{(2)}+1\right)\,,\\
g_{(2)VV} &=&
-\frac{H_{(0)}A^T_{(2)}}{V_{(0)}}\left(2+H_{(0)}V_{(0)}A^T_{(2)}\right) \,.
\end{eqnarray}
We will not need to impose the remaining equation \eqref{eq:adsleading term} in
order to make the argument\footnote{Note that only the $TT$ component of
\eqref{eq:adsleading term} would impose a further constraint.}.

By expanding the vector field equation of motion \eqref{eq:Scheom Mat} to leading
order with the fields above we find
\begin{eqnarray}
\mbox{Vec}_{T} & = & \frac{4\log^2 r}{r^2}H_{(0)}^3V_{(0)}^3(A^T_{(2,1)})^2 +
O\left(\frac{\log r}{r^2}\right) \,,
\end{eqnarray}
which forces
\begin{equation}
A^T_{(2,1)} = e^{+}_{(2,1)V} = 0\,. \label{eq: AT21 zero}
\end{equation}
Using \eqref{eq: AT21 zero} it can be seen that $\mbox{Vec}_{T}$ is of
order $r^{-2}$ and $\mbox{Vec}_{V}$ is of order $r^0$. Taking the linear combination
\begin{equation}
r^2\left(1+A^T_{(2)}H_{(0)}V_{(0)}\right)\left(\mbox{Vec}_{T}\right)+V_{(0)}^2H_{(0)}\left(\mbox{Vec}_{V}\right)
= 4H_{(0)}^2V_{(0)}^2A^T_{(2)} + \dots \,,
\end{equation}
where the dots are higher order terms we deduce that any non-vanishing $A^T_{(2)}$
is inconsistent with the equations of motion. Therefore,
\begin{equation}
A^T_{(2)} = e^{+}_{(2)V} = 0\,,
\end{equation}
and we conclude that we necessarily have the relations
\begin{eqnarray}
e^{+}_{(2,1)a}A^a_{(0)}&=&0 \,,\\
e^{+}_{(2)a}A^a_{(0)}&=&0 \,,
\end{eqnarray}
satisfied on-shell. It also follows that $\det(g_{(0)ab})=\det(\gamma_{(0)ab})$
which will allow us to introduce the $TV$ coordinates directly on $g_{(0)ab}$.

\subsection{The unconstrained case: expansions for $g_{\mu\nu}$ and $A^\mu$}\label{subsection:solvingeom}

We plug the ansatz \eqref{eq:ansatzAT}--\eqref{eq:ansatzgVV} in the equations of motion \eqref{eq:Scheom Ein}, \eqref{eq:Scheom Mat} and solve them order by order for each component. All the equations to the lowest non-vanishing order, namely $\mbox{Eins}_{TT}\; @\;r^{-4}$, $\mbox{Eins}_{TV}\; @\;r^{-2}$, $\mbox{Eins}_{VV}\; @\;r^0$, $\mbox{Eins}_{Tr}\; @\;r^{-3}$, $\mbox{Eins}_{Vr}\; @\;r^{-1}$, $\mbox{Eins}_{rr}\; @\;r^{-2}$, $\mbox{Vec}_{T}\; @\;r^{-2}$, $\mbox{Vec}_{V}\; @\;r^{0}$ and $\mbox{Vec}_{r}\; @\;r^{-1}$ are solved by
\begin{equation}
\partial_V\left(H_{(0)}V_{(0)}\right)=0\,.\label{eq: diffA0T relation}
\end{equation}

The $VV$-component of the Einstein equations at the next order is given by
\begin{eqnarray}\label{eq:Einsrrorder r-2log r}
\mbox{Eins}_{VV}\; @\;r^2\log r & = & -4g_{(4,1)VV}\,,\\
\mbox{Eins}_{VV}\; @\;r^2 & = & - \left(4g_{(4)VV}+3g_{(4,1)VV}\right)\,,
\end{eqnarray}
from which it follows that
\begin{equation}
g_{(4,1)VV}=g_{(4)VV} = 0 \,.
\end{equation}
From
\begin{equation}
\mbox{Eins}_{rr}\; @\;r^0 = -\frac{2}{H_{(0)}}g_{(2,1)TV} \,,
\end{equation}
we also find
\begin{equation}
g_{(2,1)TV} = 0 \,.
\end{equation}
The $\mbox{Eins}_{Vr}$ component reads
\begin{equation}
\mbox{Eins}_{Vr}\; @\;r = -\frac{1}{H_{(0)}V_{(0)}}\partial_V\left(g_{(2)TV}V_{(0)}\right)\,,
\end{equation}
so that
\begin{equation}\label{eq: diffg20 relation}
\partial_V\left(g_{(2)TV}V_{(0)}\right)=0\,.
\end{equation}
We next deduce from the two components
\begin{eqnarray}
\mbox{Eins}_{Tr}\; @\;r^{-1}\log r &=& -V_{(0)}\left(H_{(0)}\partial_VA^V_{(2,1)}+2H_{(0)}A^r_{(2,1)}-H^2_{(0)}V_{(0)}^2\partial_VA^T_{(4,1)}\right.\nonumber\\
&&\left.+A^V_{(2,1)}\partial_V H_{(0)}\right)\,,\label{eq: EinTr lnr^-1}
\end{eqnarray}
and
\begin{eqnarray}
\mbox{Vec}_{T}\; @\;\log r &=& V_{(0)}^2\left(-4\frac{H_{(0)}}{V_{(0)}^2}A^V_{(2,1)}-\frac{1}{H_{(0)}}A^V_{(2,1)}(\partial_V H_{(0)})^2+H_{(0)}\partial_V^2 A^V_{(2,1)} \right. \nonumber\\&& +(\partial_V A^V_{(2,1)})\partial_V H_{(0)}+H_{(0)}V_{(0)}^2(\partial_V A^T_{(4,1)})\partial_V H_{(0)}+A^V_{(2,1)}\partial_V^2 H_{(0)} \nonumber\\&& \left.+2H_{(0)}\partial_V A^r_{(2,1)}-H^2_{(0)}V_{(0)}^2\partial_V^2 A^T_{(4,1)}\right) \,,\label{eq: VecT ln}
\end{eqnarray}
that
\begin{equation}
A^V_{(2,1)} = 0\,,
\end{equation}
by plugging \eqref{eq: EinTr lnr^-1} and its $V$-derivative into \eqref{eq: VecT ln}.

Consider next the following three equations
\begin{eqnarray}
\mbox{Eins}_{TT}\; @\;r^{-2}\log r &=& \frac{1}{2}V_{(0)}^2\left(\frac{1}{H_{(0)}}(\partial_V g_{TT(0,1)}) \partial_V H_{(0)} -4H_{(0)}^2V_{(0)} \partial_VA^r_{(2,1)} \right. \nonumber\\
&& \left.+16 H_{(0)}^3 V_{(0)}A^T_{(4,1)}  - \partial^2_V g_{TT(0,1)}\right)\,,\label{eq:EinTTorder r-2log r}\\
&&\nonumber\\
\mbox{Eins}_{TV}\; @\;\log r &=& \frac{1}{2H_{(0)}}\left(\partial^2_V g_{TT(0,1)}+2 H_{(0)}^2 V_{(0)} \partial_VA^r_{(2,1)}-\frac{1}{H_{(0)}}  (\partial_V g_{TT(0,1)}) \partial_V H_{(0)}\right. \nonumber\\
&& \left. -8 H_{(0)}^3 V_{(0)} A^T_{(4,1)}\right)\,,\label{eq:EinTVorder log r}\\
&&\nonumber\\
\mbox{Eins}_{Tr}\; @\;r^{-1}\log r &=& H_{(0)} V_{(0)} \left(H_{(0)}V_{(0)}^2 \partial_V A^T_{(4,1)}-2A^r_{(2,1)}\right) \,.\label{eq:EinTrorder r-1log r}
\end{eqnarray}
Combining \eqref{eq:EinTTorder r-2log r} and \eqref{eq:EinTVorder log r} we find
\begin{eqnarray}
\partial_VA^r_{(2,1)} & = & 4 H_{(0)} A^T_{(4,1)}\,,\label{eq:partialVAr21}\\
0 & = & (\partial_V g_{TT(0,1)}) \partial_V H_{(0)} -H_{(0)} \partial^2_V g_{TT(0,1)}\,.\label{eq:partialVgTT01}
\end{eqnarray}
From equation \eqref{eq:EinTrorder r-1log r} we find
\begin{equation}
A^r_{(2,1)}=\frac{1}{2}H_{(0)}V_{(0)}^2\partial_VA_{(4,1)}^T\,.\label{eq:VAr21}
\end{equation}
It can be shown that $\mbox{Vec}_{T}$, $\mbox{Vec}_{V}$ and $\mbox{Vec}_{r}$ at orders $\log r$, $r^2\log r$ and $r\log r$, respectively, are now automatically solved.

From $\mbox{Eins}_{TT}$, $\mbox{Eins}_{TV}$ and $\mbox{Eins}_{Tr}$ at orders $r^{-2}$, $r^0$ and $r^{-1}$ we get
\begin{eqnarray}
\hspace{-.9cm} g_{TT(0,1)} & = & H_{(0)}V_{(0)}\left(2H_{(0)}V_{(0)}^2  \partial_V A^r_{(2)}+4 H_{(0)}A^V_{(2)}+4 V_{(0)} g_{(2)TV}-8H_{(0)}^2 V_{(0)}^2 A^T_{(4)}\right. \nonumber\\
&&\left.- V_{(0)} \partial_T\partial_V\log H_{(0)}\right)\,,\label{eq:EinsTTorder r-2}\\
&&\nonumber\\
\hspace{-.9cm} H_{(0)}^2 V_{(0)} A^T_{(4,1)} &=& -4H_{(0)}^2 V_{(0)} A^T_{(4)}+H_{(0)}V_{(0)} \partial_VA^r_{(2)}+ 2 g_{(2)TV}  -  \partial_T\partial_V\log H_{(0)}\,,\label{eq:EinsTVorder r^0}\\
&&\nonumber\\
\hspace{-.9cm} \partial_V g_{TT(0,1)} &=& H^2_{(0)}V_{(0)}\left(-2H_{(0)}V_{(0)}^2 \partial_V A^T_{(4)} +2H_{(0)}^{-1}A^V_{(2)} \partial_V H_{(0)} +4A^r_{(2)}+2\partial_VA^V_{(2)}\right)\,,\label{eq:EinsTrorder r-1}
\end{eqnarray}
respectively. Equations $\mbox{Vec}_{T}$ and $\mbox{Vec}_{V}$ at orders $r^0$ and $r^2$ read
\begin{eqnarray}
0 & = & 4 \frac{H_{(0)}^2}{V_{(0)}^2} A^V_{(2)}-H_{(0)} (\partial_V H_{(0)}) \partial_VA^V_{(2)}+2 H_{(0)}^3 A^T_{(4,1)}-2H_{(0)}^2 \partial^2_VA^V_{(2)}+4 \frac{H_{(0)}}{V_{(0)}} g_{(2)TV} \nonumber\\
&& +H_{(0)}^3V_{(0)}^2 \partial^2_VA^T_{(4)}+ (\partial_V H_{(0)})^2 A^V_{(2)} -2 H_{(0)}^2 \partial_VA^r_{(2)}  -H_{(0)} A^V_{(2)} \partial^2_V H_{(0)}\nonumber\\
&&-H_{(0)}^2V_{(0)}^2 (\partial_V H_{(0)}) \partial_VA^T_{(4)}\,,\label{eq:VecTorder r^0}\\
&&\nonumber\\
0 & = &  -H_{(0)}^2 \partial^2_VA^V_{(2)}-H_{(0)}^2V_{(0)}^2 (\partial_V H_{(0)}) \partial_VA^T_{(4)}-H_{(0)} (\partial_V H_{(0)}) \partial_VA^V_{(2)}-2 H_{(0)}^2 \partial_VA^r_{(2)}\nonumber\\
&& +H_{(0)}^3V_{(0)}^2 \partial^2_VA^T_{(4)} +(\partial_V H_{(0)})^2 A^V_{(2)}-H_{(0)} A^V_{(2)} \partial^2_V H_{(0)} \,,\label{eq:VecVorder r^2}
\end{eqnarray}
respectively. Combining \eqref{eq:VecTorder r^0} and \eqref{eq:EinsTVorder r^0} by eliminating $A^T_{(4,1)}$ and using \eqref{eq:VecVorder r^2} we obtain
\begin{equation}
\partial_V A^r_{(2)} = -2\frac{A^V_{(2)}}{V_{(0)}^2}+4H_{(0)}A^T_{(4)}+\frac{1}{H_{(0)}V_{(0)}}\partial_T\partial_V\log H_{(0)}-4\frac{g_{(2)TV}}{H_{(0)}V_{(0)}}\,.\label{eq:partialVAr2}
\end{equation}
Further from \eqref{eq:EinsTTorder r-2} and \eqref{eq:partialVAr2} and from \eqref{eq:EinsTVorder r^0} and \eqref{eq:partialVAr2} we get
\begin{eqnarray}
g_{TT(0,1)} & = & -4H_{(0)}V_{(0)}^2g_{(2)TV}+H_{(0)}V_{(0)}^2\partial_T\partial_V\log H_{(0)}\,,\label{eq:gTT01}\\
 A^T_{(4,1)} & = & -2\frac{g_{(2)TV}}{H_{(0)}^2V_{(0)}}-2\frac{A^V_{(2)}}{H_{(0)}V_{(0)}^2}\,,
\end{eqnarray}
respectively. With the help of \eqref{eq:gTT01} equation \eqref{eq:EinsTrorder r-1} can be written as
\begin{equation}
A^r_{(2)} = \frac{1}{2}H_{(0)}V_{(0)}^2\partial_VA^T_{(4)}-\frac{1}{2H_{(0)}}\partial_V(H_{(0)}A_{(2)}^V)+\frac{1}{4}V_{(0)}\partial_V\left(\frac{1}{H_{(0)}}\partial_T\partial_V\log H_{(0)}\right)\,.\label{eq:Ar2}
\end{equation}
Given \eqref{eq:partialVAr21}--\eqref{eq:VAr21} together with \eqref{eq:partialVAr2}--\eqref{eq:Ar2} equations $\mbox{Vec}_{T}$, $\mbox{Vec}_{V}$ and $\mbox{Vec}_{r}$ are now solved at orders $r^0$, $r^2$ and $r$, respectively, for $A^r_{(2,1)}$, $A^r_{(2)}$, $A^T_{(4,1)}$, $A^V_{(2)}$ and $g_{(0,1)TT}$ in terms of $H_{(0)}$, $V_{(0)}$, $g_{(2)TV}$ and $A^T_{(4)}$. The same is true for equations $\mbox{Eins}_{TT}$, $\mbox{Eins}_{TV}$ and $\mbox{Eins}_{Tr}$ at the respective orders $r^{-2}$, $r^0$ and $r^{-1}$.

\section{Real scalars on a fixed Schr\"odinger background}\label{app:scalars}

We show in sections \ref{eq:solvingKG} and \ref{subsec:breakdownscalar} that the scalar field on a fixed Sch\"odinger background has many features in common with the components of the metric and vector field in section \ref{sec:solutions}. We briefly discuss the scalar solutions to the free Klein--Gordon equation for a compact $V$ coordinate in section \ref{subsec:scalarcompactV}. We discuss various aspects of the scalar field theory in \ref{subsec:discussion}.

\subsection{Solving the Klein--Gordon equation}\label{eq:solvingKG}

Consider the metric of a Schr\"odinger space-time in Poincar\'e coordinates
\begin{equation}
ds^2=-\beta^2\frac{dT^2}{r^4}+\frac{1}{r^2}\left(2dTdV+dr^2\right)
\end{equation}
where $\beta^2>0$ can be scaled away. Allowing $\beta$ to be zero we can compare with the AdS case. The Klein--Gordon equation $\square\phi=0$ reads
\begin{equation}\label{eq:eomphi}
\beta^2\partial_V^2\phi+2r^2\partial_T\partial_V\phi+r^2\partial_r^2\phi-r\partial_r\phi=0\,.
\end{equation}
The most general solution is given by (see also \cite{Son:2008ye,Balasubramanian:2008dm,Barbon:2008bg,Guica:2010sw})
\begin{equation}\label{eq:scalardecomposition}
\phi=a^-_0(T)+a^+_0(T)r^2+\int d\omega dm e^{-i\omega T}e^{-imV}\phi_{\omega,m}(r)\,,
\end{equation}
where $a^-_0(T)$ and $a^+_0(T)$ are arbitrary functions of $T$ and where $\phi_{\omega,m}(r)$ satisfies
\begin{equation}
\phi^*_{-\omega,-m}=\phi_{\omega,m}\,,
\end{equation}
as well as
\begin{equation}
r^2\phi''_{\omega,m}(r)-r\phi'_{\omega,m}(r)-\beta^2 m^2\phi_{\omega,m}(r)-2m\omega r^2\phi_{\omega,m}(r)=0\,,
\end{equation}
whose general solution is
\begin{eqnarray}
\phi_{\omega,m}(r) & = & r\theta(-m\omega)\left(c^1_{\omega,m}J_\nu\left(\sqrt{-2m\omega}r\right)+c^2_{\omega,m}Y_\nu\left(\sqrt{-2m\omega}r\right)\right)\nonumber\\
&& +r\theta(m\omega)\left(c^3_{\omega,m}I_\nu\left(\sqrt{2m\omega}r\right)+c^4_{\omega,m}K_\nu\left(\sqrt{2m\omega}r\right)\right)\,.
\end{eqnarray}
In here $\theta$ is the Heaviside step function, the parameter $\nu$ is given by
\begin{equation}
\nu^2=1+\beta^2 m^2\,,
\end{equation}
and $J_\nu$ and $Y_\nu$, $I_\nu$ and $K_\nu$ are two independent Bessel and modified Bessel functions, respectively.

It follows that $\phi_{\omega,m}(r)$ can be expanded as
\begin{equation}
\phi_{\omega,m}(r)=r^{1-\nu}a^-_{\omega,m}F^-_{\omega,m}(r^2)+r^{1+\nu}\left(a^+_{\omega,m}+b_{\omega,m}\log r\right)F^+_{\omega,m}(r^2)\,,
\end{equation}
where the functions $F^-_{\omega,m}$ and $F^+_{\omega,m}$ admit a Taylor series expansion around $r=0$ such that $F^\pm_{\omega,m}(0)=1$. The coefficient $b_{\omega,m}$ is nonzero if and only if $\nu>1$ is an integer in which case $F^-_{\omega,m}$ is a polynomial of degree $2\nu-2$. The coefficients $a^-_{\omega,m}$ and $a^+_{\omega,m}$ depend on $m$ and $\omega$ and are the only two arbitrary coefficients in the expansion. If we Fourier transform back from $\omega$ to $T$ we obtain
\begin{eqnarray}
\phi & = & a^-_0(T)+a^+_0(T)r^2+\int dm e^{-imV}\left[r^{\Delta^-_m}a^-_m(T)F^-_m(T,r^2)\right.\nonumber\\
&&\left.+r^{\Delta^+_m}\left(a^+_m(T)+b_m(T)\log r\right)F^+_m(T,r^2)\right]\,,
\end{eqnarray}
where
\begin{equation}
\Delta_m^\pm=1\pm\sqrt{1+\beta^2m^2}\,,
\end{equation}
and where
\begin{equation}\label{eq:F+-}
F^\pm_m(T,r^2)=1+\sum_{k=1} f_{(k)m}^\pm(T)r^{2k}\,,
\end{equation}
with $F^-_m$ a polynomial of degree $2\nu-2$ and $b_m(T)\neq 0$ if and only if $\nu>1$ is an integer.

\subsection{Breakdown of the FG expansion}\label{subsec:breakdownscalar}

When $\beta=0$ the $\Delta_m^\pm$ are independent of $m$ so that we can do the $m$ integral order by order in the expansion leaving us with the well-known solution for a real scalar on AdS$_3$. When $\beta\neq 0$ it is no longer possible to do the integral coefficient by coefficient. The integration over $m$ changes the $r$ expansion via $\Delta^\pm_m$ in an important way. In fact what is worse, because the functions $a^-_{m}$ and $a^+_{m}$ all depend on $m$ and they are abritrary we cannot even derive what the $r$ expansion for $\phi$ as a function of $V$ is because we have no way of doing the $m$ integral, not even in principle, for the full class of solutions in a way that we can derive the $r$-expansion with coefficients depending on $T$ and $V$.

Another way to see this is to go back to equation \eqref{eq:eomphi} and naively expand it as if there is an FG type expansion. Let us therefore assume that there is a solution that can be expanded as
\begin{equation}
\phi = r^\Delta\left(\phi_{(0)} + \dots \right)\,.\label{Sch type ansatz}
\end{equation}
For example we might be interested in a solution of this form because we want to source an operator with a well-defined scaling dimension. We find from the Klein--Gordon equation that the leading order coefficient satisfies either
\begin{eqnarray}
\hspace{-.7cm}\Delta=0,2 &\;\mbox{with}\;& \phi_{(0)}=F_1(T) + F_2(T)V\,,\\
\hspace{-.7cm}0<\Delta<2 &\;\mbox{with}\;& \phi_{(0)}= F_1(T)e^{\sqrt{\Delta(2-\Delta)}V}+F_2(T)e^{-\sqrt{\Delta(2-\Delta)}V}\,,\\
\hspace{-.7cm}\Delta<0, \Delta > 2&\;\mbox{with}\;& \phi_{(0)}= F_1(T)\sin(\sqrt{\Delta(\Delta-2)}V)+F_2(T)\cos(\sqrt{\Delta(\Delta-2)}V)\,,
\end{eqnarray}
where we have set $\beta=1$. Let us choose to look at $\Delta=0$. In general it is far from clear what kind of ansatz one should take for the $r$-expansion. To gain some understanding of this problem we start with a very simple solution which is the most general solution of the KG equation $\square\phi=0$ subject to the constraint $\partial^2_V\phi=0$. We find that in this case there is a unique terminating solution given by
\begin{equation}
\phi = \phi_{(0)} + r^2\ln(r)\phi_{(21)} + r^2\phi_{(2)} + r^4\phi_{(4)}\,,
\end{equation}
where
\begin{eqnarray}
\phi_{(0)}&=&F_1(T) + F_2(T)V\,,\\
\phi_{(2)}&=&F_3(T) + F_4(T)V\,,\label{phi2 Sch simp}\\
\phi_{(21)}&=&-\frac{d}{dT}F_2(T)\,,\\
\phi_{(4)}&=&-\frac{1}{4}\frac{d}{dT}F_4(T)\,.
\end{eqnarray}

Having this result we now want to extend the solution space. The problem we encounter is that the solution we will typically get by solving $\square\phi=0$ with a FG type ansatz like \eqref{Sch type ansatz} depends on the ansatz we make. In particular, if we assume an infinite series for $\phi$ of the form\footnote{We do not exclude the possibility that a more general ansatz for the expansion of $\phi$ contains additional terms that are not of the form $r^{2n}\log^k r$.}
\begin{equation}
\phi=\phi_{(0)}+\sum_{n,k} r^{2n}\log^k r\phi_{(2n,k)}\,,
\end{equation}
there seems to be no restriction on the powers of the logs. More precisely, having a term of order $r^2\log r$ in the ansatz we find that in general $\phi_{(2)}$ (we use the notation $\phi_{(2n,0)}=\phi_{(2n)}$) is not given by \eqref{phi2 Sch simp} only but has additionally terms proportional to $V^2$ and $V^3$. Allowing for a term of order $r^2\log^n r$ in the ansatz we will find that the $V$-dependence in $\phi_{(2)}$ is at most of order $V^{2n+1}$. So the more powers of $\log r$ we add to the ansatz the less constrained $\phi_{(2)}$ will be (the more powers of $V$ we can include).

For explicitness and in order to analyze other questions we now fix the ansatz by hand to be of the form
\begin{equation}\label{scalaransatz}
\phi=\phi_{(0)}+\sum_{n=1}^\infty r^{2n}\sum_{k=0}^n\log^k r\phi_{(2n,k)}\,.
\end{equation}
Up to NNLO the solution can be written as
\begin{eqnarray}
\phi_{(0)}&=& F_1(T) + VF_2(T)\,,\\
\phi_{(2)}&=&F_3(T) + VF_4(T) + V^2F_5(T) + V^3F_6(T)  \,,\\
\phi_{(2,1)}&=&-F'_2(T) - F_5(T) - 3VF_6(T)\,,\\
\phi_{(4)}&=&-\frac{1}{4}F'_4(T)-\frac{1}{2}VF'_5(T)-\frac{3}{8}\left(1+2V^2\right)F'_6(T)+\varphi_{(4)}\,,\\
\phi_{(4,1)}&=& \frac{3}{4}F'_6(T)-\frac{1}{216}\left(\partial^2_V+8\right)\left(\partial^2_V+44\right)\varphi_{(4)}\,,\\
\phi_{(4,2)}&=& \frac{1}{72}\left(\partial^2_V+8\right)^2\varphi_{(4)}\,,
\end{eqnarray}
where $\varphi_{(4)}$ has to satisfy the differential equation
\begin{equation}
\left(\partial^2_V+8\right)^3\varphi_{(4)}= 0\,.
\end{equation}
Solving this differential equation leads to new arbitrary functions of $T$ that appear as integration constants. At higher orders we find that all log terms $\phi_{(2n,1)}$ are determined in terms of the $\phi_{(2n)}$. However, each $\phi_{(2n)}$ for $n>2$ contains new undetermined functions in a similar fashion to what happened at order $n=2$.

We thus clearly see that there is no FG expansion as new functions keep appearing at higher orders. On top of that we have no control over the ansatz as we can keep adding arbitrarily high powers of $\log r$ at each order in $r$. When we solve the equations of motion of the massive vector model in the radial $TV$ gauge in section \ref{sec:solutions} it can be observed that the components of the vector field will behave in a similar way to the real scalar field discussed here.

\subsection{Compactifying the $V$ coordinate}\label{subsec:scalarcompactV}

If we compactify $V\sim V+2\pi L$ the solution to the Klein--Gordon equation $\square\phi=0$ becomes
\begin{equation}
\phi = \sum_{m\in\mathbb{Z}} e^{-imV/L}\left[r^{\Delta^-_m}a^-_m(T)F^-_m(T,r^2)+r^{\Delta^+_m}\left(a^+_m(T)+b_m(T)\log r\right)F^+_m(T,r^2)\right]\,,
\end{equation}
where the functions $F^\pm_m$ are of the form \eqref{eq:F+-} and where
\begin{equation}
\Delta^\pm_m=1\pm\sqrt{1+\frac{\beta^2m^2}{L^2}}\,,
\end{equation}
with the usual specifications whenever $\nu=\sqrt{1+\frac{\beta^2m^2}{L^2}}\in\mathbb{N}$. This means that whenever $\frac{\beta^2}{L^2}\in\mathbb{Q}$ there will always be a subset of the $m\in\mathbb{Z}$ for which $\nu\in\mathbb{N}$ and that whenever $\frac{\beta^2}{L^2}\notin\mathbb{Q}$ we have that $\nu\notin\mathbb{N}$ for all values of $m$. The functions $a^-_0(T)$ and $a^+_0(T)$ now correspond to the $m=0$ modes. If we for simplicity assume that $\frac{\beta^2}{L^2}\notin\mathbb{Q}$ then we have
\begin{eqnarray}
\phi & = & a^-_0(T)+a^+_0(T)r^2\nonumber\\
&&+e^{-iV/L}\left[r^{\Delta^-_1}a^-_1(T)F^-_1(T,r^2)+r^{\Delta^+_1}a^+_1(T)F^+_1(T,r^2)\right]+\text{c.c.}\nonumber\\
&&+e^{-2iV/L}\left[r^{\Delta^-_2}a^-_2(T)F^-_2(T,r^2)+r^{\Delta^+_2}a^+_2(T)F^+_2(T,r^2)\right]+\text{c.c.}\nonumber\\
&&+\ldots\,.\label{eq:phiVnoncompact}
\end{eqnarray}

\subsection{Discussion}\label{subsec:discussion}

On AdS we consider the most general solution to the Klein--Gordon equation because these correspond to unitary irreducible representations (UIRs) of the AdS isometry group and are therefore needed to compute correlation functions of scalar operators in the dual CFT. On a Schr\"odinger space-time this is totally different. The most general solution is reducible and can be decomposed into UIRs of the Schr\"odinger group. The scalar UIRs of the Schr\"odinger group are of the form $e^{-imV}\psi_m(T,r)$ with $m\neq 0$ that form eigenstates of the particle number generator $i\partial_V$. We thus see that each $m$-mode in \eqref{eq:scalardecomposition} or \eqref{eq:phiVnoncompact} forms a UIR of the Schr\"odinger group. Hence for the purpose of computing correlation functions of say chiral primaries as defined in \cite{Nishida:2007pj} we do not need more general solutions.

Further in \cite{Blau:2010fh} it has been shown that despite the absence of a time function (because the Schr\"odinger space-time is non-distinguishing as discussed in \cite{Hubeny:2005qu,Blau:2010fh}) there is at least a well defined field theory for scalar fields on a Schr\"odinger space-time that form UIRs. In fact for the free theory the momentum $m$ plays the role of the Bargmann superselection parameter \cite{Bargmann:1954gh} that labels the different Hilbert spaces. By this we mean that in the free theory unitary time evolution of states of the form $e^{-imV}\psi_m(T,r)$ cannot change the value of $m$.

Nevertheless one can readily write down solutions of the free theory that are not in an eigenstate of particle number. By Fourier transform these can always be viewed as a superposition of the states of different $m$. It would be interesting to understand better the physical nature of such states and how we should think of them as states in a Hilbert space.

\bibliographystyle{newutphys}
\bibliography{Schroedinger}

\end{document}